\documentclass[reqno,a4paper]{amsart}
\usepackage{amsmath,paralist,amssymb,amsfonts,epsfig,float,color,cite,comment,subfigure}
\definecolor{lightblue}{rgb}{0.22,0.45,0.70}
\definecolor{cgray}{rgb}{0.9,0.9,0.9}
\usepackage{dsfont}
\usepackage[colorlinks=true,breaklinks=true,linkcolor=lightblue,citecolor=lightblue]{hyperref}

\def\L{\mathrm{L}}

\newcommand{\Grad}{\nabla}

\numberwithin{equation}{section}

\renewcommand\L{\mathrm{L}}

\renewcommand\L{\mathrm{L}}

\newtheorem{theorem}{Theorem}[section]
\newtheorem{lemma}[theorem]{Lemma}

\newtheorem{proposition}{Proposition}[section]

\usepackage{tikz}
\usepackage{pgfplots}
\usetikzlibrary{spy,calc}

\title[]{A nonlinear cross-diffusion epidemic with time-dependent SIRD system: Multiscale derivation and computational analysis}
\author[]{Mohamed Zagour}

\begin{document}
\maketitle



\begin{abstract}
	
 A nonlinear cross-diffusion epidemic with a time-dependent Susceptible-Infected-Recovered-Died system is proposed in this paper. This system is derived from kinetic theory model by multiscale approach, which leads to an equivalent system coupled the microscopic and macroscopic equations. Subsequently, numerical investigations to design asymptotic preserving scheme property is developed and validated by various numerical tests. Finally, the numerical computational results of the proposed system are discussed in two dimensional space using the finite volume method.
\end{abstract}

\date{\today }

\section{Introduction} \label{Intro}
The outbreak of the new coronavirus, called COVID-19, caused by severe acute respiratory syndrome coronavirus 2 (SARS-CoV-2) appeared,  in December 2019, apparently occurred in Wuhan, China. The spread of the epidemics has been very fast this covering all countries in the world. Thus, pandemic has severely affected the economy, health, and security of the society all over the world. Data are so impressive, as by August 2021 more than 200 million  people have been  infected and more than 4 million people died \cite{[Rf1]}.

As it is known, mathematical models may help  decision making, for example, about containment measures, lock-down, and vaccination campaigns. Indeed, they can contribute both to research in epidemiology and to crisis managers, however without naively claiming that mathematics can tackle the problem of derivation of models by a standing alone approach. For instance, models can depict a variety of epidemic scenarios. In addition, they can contribute to a deeper  understanding the contagion mechanisms. 

Several models have been proposed in the literature to describe the dynamic of epidemics which can be classified as network or collective  models. The first class treat a population as a network of interacting individuals, and the contagion process is described at the microscopic scale see~\cite{[W14]}. Data on the spread of the epidemics are available in~\cite{[Fox20],[LZ20],[PLZ20]}. Modeling of vaccination dynamics and medical actions are treated in~\cite{[LT20],[SW20]}.

 Collective models describe the spread of the epidemic in a population using a limited number of collective variables with a small number of parameters. For instance, celebrated logistic models~\cite{[KM27],[V38]}, Richards models~\cite{[R59]}, susceptible-infected-recovered (SIR)  models \cite{[B75],[KM27]}, and  susceptible-exposed-infectious-removed (SEIR) models~\cite{[C17]}. 

It is worth to mention that the classical SIR, SEIR, and other similar models belong to the class of compartmental models \cite{[B17],[C17],[R08],[SZ20]}. An exhaustive presentation, which includes qualitative analysis and biological applications can be found in \cite{[PERTH]}. However, the the paper \cite{[Est2020]} introduces  conceivable derivation of theses models as natural development of classical SEIR models. The interested reader is addressed to~\cite{[BB20],[Est2020]} not only for a broad reference to the existing literature, but also for various challenging research perspectives.

This paper is devoted to a multiscale derivation approach  of  time-dependent nonlinear SIRD cross-diffusion system (\ref{Cross-Diffusion}) from kinetic theory model by using the micro-macro decomposition method. Firstly, the kinetic theory model is rewritten as coupled system of microscopic part and macroscopic one and subsequently macroscopic models are derived by low order asymptotic expansions in terms of a small parameter. Note that this approach has been applied to the micro-macro application in different fields. For example, a time-dependent SEIRD reaction diffusion \cite{[Z21]}, chemotaxis phenomena related to Keller-Segel model \cite{[BBNS12]}, and formation of patterns induced by cross-diffusion in a fluid~\cite{[ABKMZ19],[BKZ18]}. This technique motivated the design  numerical tools that preserve the asymptotic property \cite{[JI99],[KL98]}. Concretely, these methods design the uniform stability and consistency of numerical schemes in the limit along the transition from kinetic to macroscopic regimes.

Motivated by the obtained numerical results in one dimensional space, this paper is also deals with the computational analysis in two dimensional space using finite volume method. We provide the pattern formation induced by cross-diffusion term.  In the modeling point of view, the term cross-diffusion has the interpretation that the susceptible SP moves away from the increasing gradients of the infected SP. In addition, it is assumed that the cross-diffusion effect depends on the local population density. Thus, for nonlinear cross-scattering, recklessness exists at a small number and fatalism at a high total population number. With recklessness and fatalism, the susceptible subpopulation decreases its tendency to avoid the agents of the infected population.

The rest of this paper is organized as follows: in Section \ref{Sec2}  we present a phenomenological derivation of a macro-scale model of virus contagion and cross-diffusion in space. Section \ref{Sec3} briefly presents the multiscale approach by micro-macro method which leads to the derivation of system (\ref{Cross-Diffusion}) from a kinetic theory model. Section \ref{Sec4} is devoted to the development of an asymptotic preserving numerical scheme in one dimensional space by finite volume method. The aim is to guarantee the  uniform stability with respect to Knudsen parameter $\varepsilon$, related to the mean distance between individuals, as well as consistency with the cross-diffusion limit. In addition, we provide some numerical simulations obtained with the equivalent micro-macro formulation and also with the macroscopic scheme, where we show the asymptotic preserving scheme property. In addition, we show the role of presence of the diffusion terms in system (\ref{Cross-Diffusion}), and its sensitivity with respect to the different choices of the reproduction ratio $R_0$. Finally, motivated by the obtained numerical results in one dimensional space, Section \ref{Sec5} provide numerical results in two dimensional space using finite volume method of formation of patterns.
\section{Phenomenological modeling of nonlinear cross-diffusion population dynamics}\label{Sec2}
We consider a population constituted by $N_0$ individuals which can be subdivided into a number of sub-population, in short SP, each characterized by a different biological state. Specifically, we consider the following SP whose states are defined by their number, referred to $N_0$, depending on time and space, where individuals correspond to:
\begin{enumerate}
	
	\vskip.1cm \item $N(t,x)$ Alive;
	
	\vskip.1cm \item $S(t,x)$ Susceptible;
	
	\vskip.1cm \item $I(t,x)$ Infected;
	
	\vskip.1cm \item $R(t,x)$ Recovered;
	
	\vskip.1cm \item $D(t,x)$ Died.
	
\end{enumerate}

Accordingly, the aforementioned normalization with respect to $N_0$ implies that
$$
N(t,x) + S(t,x)+ I(t,x) + R(t,x) + D(t,x) = 1, \hskip1cm t \geq 0, \hskip1cm x \in \Omega,
$$
where $\Omega$ is a bounded domain within which the population is confined.

The multiscale derivation of our proposed macroscopic system can be obtained according to the following assumptions:

\begin{enumerate}
	
	\item Individuals diffuse within the domain $\Omega$ by a nonlinear diffusion function $\varphi(x,h)$ depending on a spacial distribution which considers the preferred directions of propagation and the on density of SP;
	
	\item The susceptible SP moves away from increasing gradients of the infected SP. This can be modeled by a cross-diffusion term;
	
	\item The interaction dynamics is modeled by a source term involving the interactions of different SPs;
	
	\item Modeling of interactions accounts uses the parameters reported in Table \ref{tab2} which also reports the parameters underlying the assumptions of the interaction dynamics;
	
	\item Susceptible SP may become infected due to contact with infectious individuals with a transmission rate function $ \beta(t)$, while infectious SP recovers with a $ \gamma $ rate;
	
	\item The time-dependent transmission rate function $\beta(t)$ incorporates the impact of mandatory government actions (i.e total or partial lockdown), respecting sanitary protocol and vaccination campaigns. 
	
\end{enumerate}

\begin{table}[h!]
	\begin{center}
		\caption{Description of the parameters of the SIRD system with vital dynamics and constant population}
		\label{tab2}
		\begin{tabular}{|c||l|} 
			\hline
			\textbf{Parameter} & \textbf{Description} \\
			\hline\hline
			$A$ & Recruitment rate assumed $A=\mu\,N$\\
			\hline
			$\mu$ & Natural death rate for susceptible individuals\\
			\hline
			$\beta(t)$ & Transmission rate function\\
			\hline
			$\gamma$ & Recovery rate of infectious individuals  \\
			\hline
			
		\end{tabular}
	\end{center}
\end{table}
Assumptions (1)-(6), by straightforward calculations, yield the following nonlinear cross-diffusion SIRD system with vital dynamics and constant population:
\begin{equation}\label{Cross-Diffusion}
\begin{cases}
\displaystyle \partial_t S=d_1\nabla \cdot(\varphi_1(x,S) \nabla S)+ \nabla \bigl(\chi(S,I)\Grad
I\bigr)+A-\mu S-\beta(t)S\frac{I}{N},\\
{}\\
\displaystyle \partial_t I=d_2\nabla \cdot(\varphi_2(x,I) \nabla I)+\beta(t)S\frac{I}{N}-(\mu +\gamma) I,
\\
{}\\
\displaystyle \partial_t R=d_3\nabla \cdot(\varphi_3(x,R) \nabla R)+\gamma I-\mu R, \\
{}\\
\displaystyle \partial_t D=\alpha I,
\end{cases}
\end{equation}
where $ d_i, \, i = 1, 2, 3 $ are the self-diffusion coefficients considered positive constants.

\noindent Mathematical model \ref{Cross-Diffusion} is implemented with the following initial and boundary conditions:
\begin{equation}\label{Problem}
\left\{\begin{array}{ll}
\displaystyle\frac{\partial S}{\partial \nu}=\frac{\partial I}{\partial \nu}=\frac{\partial R}{\partial \nu}=0,& x\in \partial\Omega,\; t>0,\\\\
\displaystyle S(0,x)=S_{0}(x),\;I(0,x)=I_{0}(x),& x\in \Omega,\\\\
\displaystyle R(0,x)=R_{0}(x),\;D(0,x)=D_{0}(x), &x\in \Omega.
\end{array}
\right.
\end{equation}

Note that if $ \chi = 0 $ and $\varphi_i=1$, system (\ref{Cross-Diffusion}) reduces to the reaction-diffusion SIR system \cite{[ABLN08],[CC10],[LZ11],[WLW15]}. For instance, the authors in \cite{[ABLN08]} provide a qualitative analysis to explore the impact of spatial heterogeneity of environment and human movement on the persistence and extinction of a disease. While, the authors in \cite{[CC10]} investigate analytically and numerically the behavior of positive solutions to a spatial SIR reaction–diffusion model.

Considering a time dependent transmission function can help to well model the different strategies taken to defeat the virus, for instance partial or total lockdown and the vaccination campaign. We mention that the basic reproduction ratio, denoted by $R_0$, is the classical epidemiological measure associated with the reproductive power of the disease. It is used to estimate the growth of the viral epidemic. For our system (\ref{Cross-Diffusion}) it is given by the following function
\begin{equation}
R_0(t)=\frac{\beta(t)}{\gamma+\mu},
\end{equation}
which provides a threshold for disease-free equilibrium point stability. Indeed, if $R_0(t) < 1$, the disease goes out; while if $R_0(t)> 1$, an epidemic occurs, see e.g. \cite{[H00]}.

Recently, the author in \cite{[Z21]} proposed a time-dependent SEIRD reaction-diffusion model with the following features : $i)$ a transmission rate function rather than a constant and $ii)$ the diffusion of individuals depends on a spatial distribution which considers the preferred directions of propagation modeled by a coefficient which models the diffusion coefficient in the territory. Specifically, the model takes into account both transport and diffusion and, subsequently, the modeling of these terms takes into account the specific geography of the territory and, in particular, the transport network. However, this subject has been developed in \cite{[BRR21],[BP21],[BDP21]}.  In this paper, basing on the aforesaid paper, we proposed an improved model which takes into account the nonlinear self-diffusion depending on the density of the SP, namely $\varphi_i(x,h)$. In addition, we add the cross-diffusion term  $\nabla \dot(\chi(S, I)\nabla I) $ in the dynamic of the susceptible SP. Indeed, this allows the susceptible SP to avoid the infected SP by the added cross-diffusion term. Concretely, the cross-diffusion term directs the flow in the opposite direction of the gradient $\nabla I$ whenever there is an increase of the amount of  the infected SP, consequently the susceptible SP moves away from the direction of the increasing gradient \cite{[BR11],[SLL09]}.

\section{From kinetic theory model to  SIRD cross-diffusion system}\label{Sec3}
This sections deals with a multiscale approach to derivation of the time-dependent SIRD cross-diffusion system \eqref{Cross-Diffusion} from  kinetic theory model on the basis of the  micro-macro decomposition technique. We start with presenting the properties of the kinetic theory model. Then, we rewrite it as coupled system of microscopic part and macroscopic one. Finally, we derive macroscopic models by low order asymptotic expansions in terms of a small parameter $\varepsilon$ that measures the distance between individuals. 
\subsection{Kinetic theory model}
The kinetic theory model can be stated adopting the parabolic-parabolic scaling limit as follows  for $i=1,2,3$
\begin{equation}\label{so}
\left\{
\begin{array}{l}
\displaystyle\varepsilon \partial_tf_i+ v  \cdot \nabla_{x} f_i=
\frac{1}{\varepsilon}\mathcal{T}_i[f_1,\cdots,f_{i-1},f_{i+1},\cdots,f_3](f_i) +\varepsilon\, G_i(f_1,\dots,f_3),    
\\
{}\\
\displaystyle \partial_t D=\alpha \int_Vf_2\,dv,\\\\
f_i(0,x,v)=f_{i,0}(x,v),\; D(0,x)=D_0(x),
\end{array}\right.
\end{equation}
where $f_1(t,x,v),\,f_2(t,x,v),\, f_3(t,x,v)$ are the distribution functions describing the statistical evolution of susceptible, infected and recovered individuals, respectively. $t>0$, $ x\in \mathbb{R}^{d}$, $v \in V$ are respectively, time, position and velocity. The term  $\mathcal{T}_i$ is the stochastic operator representing a random modification of direction of individuals and the operator $G_i$ ($i=1,2,3$) describes their gain-loss balance.. \\
The micro-macro decomposition technique is based on the following assumptions. \\
\textbf{Assumption 1:} The turning operator $\mathcal{T}_i$ is decomposed as follows:
\begin{equation}
\mathcal{T}_i[f_1,\cdots,f_{i-1},f_{i+1},\cdots,f_3](f_i)= \mathcal{L}_i(f_i)+\varepsilon\,\mathcal{T}_i^2[f_1,\cdots,f_{i-1},f_{i+1},\cdots,f_3](f_i),
\end{equation}
where $\mathcal{L}_i$ represents the dominant part of the turning kernel and is assumed to be independent of $f_1,\cdots,f_{i-1},f_{i+1},\cdots,f_3$. The operators $\mathcal{T}_{i}^j$ for $i=1,2,3$ and $j=1,2$ are given by
\begin{equation}
\displaystyle\mathcal{T}_{i}^j(f_i)= \int_{V}\big(T_{i}^j(v^*,v)f_i(t, x, v ^{*}) -
T_{i}^j(v,v^*) f_i(t, x, v ) \big)dv ^{*}, \label{3.4}
\end{equation}
where $T_i^j$ is the probability kernel for the new velocity $v \in V$ given that the previous velocity was $v^*$. \\
\textbf{Assumption 2:} We assume that the operators $\mathcal{T}_i$ satisfy
\begin{equation}\label{H0}
\displaystyle \int_V \mathcal{T}_i\,dv =\int_V \mathcal{L}_i\,dv=\int_V \mathcal{T}_i^2\,dv=0, \;\;i=1,2 ,3,
\end{equation}
and that there exists a bounded velocity distribution $M_i(v)>0$ independent of
$t$ and $x$ such that
\begin{equation}\label{tx}
T_i^1(v ,v ^{*} ) M_i(v ^{*}) =  T_i^1 (v ^{*},v  ) M_i(v),
\end{equation}
holds. \\
\textbf{Assumption 3:} The flow produced by these equilibrium
distributions vanish and $M_i$ are normalized, i.e.
\begin{equation}\int_V v  \, M_i(v )dv   =0, \quad \int_V
M_i(v ) dv  =1, \quad  i=1,2,3. \label{equilibre}
\end{equation}
Regarding the probability kernels, we assume that $T_i^1(v ,v ^{*})$ is bounded,
and there exist a constant $\sigma_i >0$ ($ i=1,2,3$),
such that
\begin{eqnarray}\label{cx}
T_i^1(v ,v ^{*})\geq \sigma_i M_i(v ),
\end{eqnarray}
for all $ (v ,v ^{*}) \in V\times V $,  $ x \in \Omega $ and $t>0$.

\noindent Using the same arguments as in \cite{[ABKMZ19]}, the operator $\mathcal{T}_i$ has the following properties.
\begin{lemma}
	\label{LE1} If \textbf{Assumptions 1-2-3} are satisfied. Then, the following properties of the operator $\mathcal{T}_i$ for $i=1,2,3$ holds true
	\begin{itemize}\label{L1}
		\item[i)] The operator $\mathcal{L}_i$ is self-adjoint in the space
		$\displaystyle{{\L^{2}\left(V ,{dv \over M_i(v)}\right)}}$. 
		\item[ii)]
		For $f\in \L^2$, the equation $\mathcal{L}_i(g) =f$ has a unique
		solution $\displaystyle{g \in \L^{2}\left(V,\frac{dv}{
				M_i(v)}\right)}$, satisfying
		$$\int_Vg(v) dv  = 0 \quad \Longleftrightarrow \quad   \int_{V} f(v )\, dv  =0. 
		$$
		\item[iii)]  The equation $\mathcal{L}_i(g) =v  \,  M_i(v)$, has a
		unique solution denoted by $\theta_i(v )$ for $ i=1,2,3$. 
		\item[iv)]  The kernel
		of $\mathcal{L}_i$ is $N(\mathcal{L}_i) = vect(M_i(v))$ for $ i=1,\dots      ,3$.
	\end{itemize}
\end{lemma}
\subsection{The equivalent micro-macro formulation}
Here we rewrite the kinetic theory model (\ref{so}) as a coupled system of microscopic part and macroscopic one. 
\noindent We decompose the distribution function $f_i$ for $i=1,2,3$ as follows
$$f_i(t,x,v )=M_i(v) u_i(t,x) + \varepsilon  g_i(t,x,v), $$
where 
$$u_i(t,x)= \langle f_i(t,x,v )\rangle:=\int_Vf_i(t,x,v )\,dv .$$ 
Thus, $\langle g_i \rangle=  0$ for $ i=1,2,3$. Inserting $f_i$ in the kinetic theory model (\ref{so}) and using the above stated assumptions and properties of the turning operators, one has 
\begin{equation}\label{n} 
\left\{
\begin{array}{l l}
\displaystyle\partial_t (M_i (v)u_i)  + \varepsilon \partial_t g_i +
\frac{1}{\varepsilon}  v  M_i(v ) \cdot \nabla u_i + v  \cdot
\nabla g_i   = \frac{1}{\varepsilon}\mathcal{L}_i(g_i)\\\\
\hskip1cm\displaystyle+\frac{1}{\varepsilon}\mathcal{T}_i^2[f_1,\cdots,f_{i-1},f_{i+1},\cdots,f_3](M_iu_i)+\mathcal{T}_i^2[f_1,\cdots,f_{i-1},f_{i+1},\cdots,f_3](g_i)\\\\
\hskip1cm+G_{i}(f_1,f_2,f_3 )\\
{}\\
\displaystyle \partial_t D=\alpha u_2.
\end{array} \right.
\end{equation}

\noindent In order to separate the macroscopic density $u_i(t,x)$ and microscopic quantity $g_i(t,x,v )$ for $i=1,2,3$, we use the projection technique. For that, we consider $P_{M_i}$ the orthogonal projection onto $N(\mathcal{T}_i)$, for $i=1,2,3$. It follows
$$P_{M_i(v)}(h)= \langle h\rangle M_i(v), \quad  \mbox{for any}\quad  h\in
\displaystyle{{\L^{2}\left(V ,{ dv  \over M_i(v )}\right)}}, \qquad i=1,2,3.$$ 
Consequently, inserting the operators $I-P_{M_i}$ into Eq. (\ref{n}), using known properties for the projection
$P_{M_i}\;i= 1,2,3$ and integrating this equation with respect to the variable $v$ yields the equivalent micro-macro formulation
\begin{equation}\label{mM1}
\left\{
\begin{array}{l l}
\displaystyle\partial_t g_i +
\frac{1}{\varepsilon^2} v  M_i(v) \cdot \nabla u_i+ \frac{1}{\varepsilon}(I-P_{M_i})(v 
\cdot\nabla g_i)=\frac{1}{\varepsilon^2}\mathcal{L}_i(g_i)
\\\\
\hskip1cm\displaystyle+\frac{1}{\varepsilon}\mathcal{T}_i^2[f_1,\cdots,f_{i-1},f_{i+1},\cdots,f_3](M_iu_i)+\mathcal{T}_i^2[f_1,\cdots,f_{i-1},f_{i+1},\cdots,f_3](g_i)\\ \\  
\displaystyle\hskip1cm
+ \frac{1}{\varepsilon}(I-P_{M_i}) G_{i}(f_1,\dots  ,f_3), \\
{}\\
\displaystyle\partial_t u_i+ \langle v    \cdot
\nabla g_i \rangle =  \langle G_{i}(f_1,f_2,f_3)\rangle
\\
{}\\
\displaystyle \partial_t D=\alpha u_2.
\end{array} \right.
\end{equation}

The micro-macro formulation \eqref{mM1} is equivalent to kinetic model (\ref{so}) thanks to the following proposition
\begin{proposition}	
	\noindent i) Let $\displaystyle(f_1,f_2,f_3)$ be a solution of kinetic theory model (\ref{so}). Then \\$\displaystyle(u_1,u_2,u_3,g_1,g_2,g_3)$  is a solution of micro-macro formulation
	(\ref{mM1}) associated with the following initial data for $i=1,2,3$
	\begin{equation} \label{er4} \displaystyle u_i(t=0)=u_{i,0} =\langle f_{i,0} \rangle, \quad
	g_i(t=0)=g_{i,0}={1\over \varepsilon}(f_{i,0}-M_i u_{i,0}). 
	\end{equation}	
	\noindent ii) Conversely, if $\displaystyle (u_1,u_2 ,u_3,g_1,g_2,g_3)$ is a solution of micro-macro formulation
	(\ref{mM1}) associated with the following initial data $\displaystyle(u_{1,0},\dots,u_{3,0}, g_{1,0},\dots    ,g_{3,0})$ such
	that $\displaystyle\langle g_{i,0} \rangle=0$. Then $\displaystyle (f_1,f_2,f_3)$  is a solution of the kinetic model (\ref{so}) with initial data
	$\displaystyle f_{i,0}=M_i u_{i,0}+\varepsilon g_{i,0}$ and we have $ u_i=\langle f_i \rangle$ and $\langle g_i\rangle=0$, for $i=1,2,3$.
\end{proposition} 

\noindent Now, to develop asymptotic analysis of
the equivalent micro-macro formulation (\ref{mM1}), the interacting operators $\mathcal{T}_i^2$ and $G_{i}$ are assumed to satisfy the following asymptotic behavior in the limit
$$\displaystyle \mathcal{T}_i^2[M_1u_1 +\varepsilon g_1,\dots,M_{i-1}u_{i-1}+\varepsilon g_{i-1}, M_{i+1} u_{i+1}+\varepsilon g_{i+1},\dots,M_3u_3+\varepsilon g_3]$$
\begin{equation} \label{T} 
\begin{array}{l}
= \mathcal{T}_i^2[M_1u_1,\dots, M_{i-1}u_{i-1},M_{i+1}u_{i+1},\dots,M_3u_3 ]+ O(\varepsilon),
\end{array}
\end{equation}
and
\begin{equation} \label{G} 
	\displaystyle G_{i}\Big(M_1(v )u_1 +\varepsilon g_1,\dots , M_3(v )u_3 +\varepsilon g_3\Big)= G_{i}\Big(M_1(v)u_1,\dots, M_3(v )u_3 \Big)+ O(\varepsilon), 
\end{equation}
for $ i=1,2,3$.
One can obtain a general macroscopic model as $\varepsilon$ goes to $0$ from the equivalent micro-macro formulation (\ref{mM1}). Indeed, using (\ref{G}) and (\ref{mM1}), one has for $i=1,\dots,3$
$$\mathcal{L}_i(g_i) =  v  M_i(v) \cdot \nabla  u_i-\mathcal{T}_i^2[M_1u_1,\dots,M_{i-1}u_{i-1}, M_{i+1}u_{i+1},\dots,M_3u_3 ](M_iu_i)$$

\noindent From Lemma \ref{L1}, property $ii)$, the operator $\mathcal{T}_i$ is invertible. This implies 
\begin{eqnarray}\label{x1} g_i =  \mathcal{L}_i^{-1}\Big(v  M_i(v) \cdot \nabla u_i-\mathcal{T}_i^2[M_1u_1,\dots,M_{i-1}u_{i-1}, M_{i+1}u_{i+1},\dots,M_3u_3 ](M_iu_i)\Big)+O(\varepsilon).
\end{eqnarray}
Inserting (\ref{x1}) into the second equation in (\ref{mM1}) yields the following macroscopic system
\begin{equation}\label{G1}
\begin{array}{l l}
\displaystyle \partial_t u_i +  \Big\langle v  \cdot \nabla \mathcal{L}_i^{-1}\Big(v  M_i(v) \cdot \nabla  u_i-\mathcal{T}_i^2[M_1u_1,\dots,M_{i-1}u_{i-1}, M_{i+1}u_{i+1},\dots,M_3u_3 ](M_iu_i)\Big)\Big\rangle \\\\
\displaystyle\hskip5cm= \Big\langle G_{i}(M_1(v ) u_1,M_2(v )u_2, M_3(v )u_3)\Big\rangle + O(\varepsilon).
\end{array} 
\end{equation}
Thanks to the following equalities
$$\displaystyle\left\langle v \cdot \nabla
\mathcal{L}_i^{-1}\Big(v  M_i(v ) \cdot \nabla  u_i\Big)\right\rangle= \nabla\cdot\Big( \left\langle v  \otimes
\theta_i(v )\right\rangle\cdot \nabla u_i \Big), $$
and
\begin{equation*}
\begin{array}{l l}\displaystyle\left\langle v \cdot \nabla
\mathcal{L}_i^{-1}\Big(\mathcal{T}_i^2[M_1u_1,\dots,M_{i-1}u_{i-1}, M_{i+1}u_{i+1},\dots,M_3u_3](M_iu_i)\Big)\right\rangle\\
= \nabla\cdot\Big<\frac{\theta_i(v)}{M_i(v)}u_i\mathcal{T}_i^2[M_1u_1,\dots,M_{i-1}u_{i-1}, M_{i+1}u_{i+1},\dots,M_3u_3](M_i)\Big>
\end{array} 
\end{equation*} 
where $\theta_i(v)$ are given in Lemma \ref{LE1} for $ i=1,2,3$,
one has the following general macroscopic system
\begin{equation}\label{mM2}
\left\{
\begin{array}{l l}
\partial_t u_i + \nabla\cdot\Big(\Gamma_i\big(u_1,\dots,u_{i-1}, u_{i+1},\dots,u_3\big)u_i-D_i \cdot \nabla u_i\Big)=H_i(u_1,\dots    ,u_3) +O(\varepsilon),\
\\\\
\displaystyle\partial_t D=\alpha u_3,
\end{array}
\right.
\end{equation}
\noindent where $D_i$ and the functions $\Gamma_i$, $H_i$ are given by
\begin{equation}\label{di}
\qquad D_i=- \big<v  \otimes \theta_i(v) \big>  , 
\end{equation}
\begin{equation}\label{Gamma}
\Gamma_i=-\Big<\frac{\theta_i(v)}{M_i(v)}u_i\mathcal{T}_i^2[M_1u_1,\dots,M_{i-1}u_{i-1}, M_{i+1}u_{i+1},\dots,M_3u_3](M_i)\Big>,
\end{equation}
\begin{equation} \label{H}H_i(u_1,\dots,u_3)= \Big<G_{i}(M_1(v) u_1,\dots    , M_3(v) u_3) \Big>  ,\;  \hbox{for}\; i=1,2,3. \end{equation}
To derive system (\ref{Cross-Diffusion}) we consider specific choices in (\ref{so}) of the terms that appeared in the kinetic model (\ref{so}). Namely
$$u_1=S, \quad u_2=I,\quad u_3=R. $$
The probability kernel $T_i$ is given by 
\begin{equation*}
T_i^1= \frac{\sigma_i}{M_i(v)},\quad \hbox{for} \; i=1,2,3.
\end{equation*} 
This implies  
\begin{equation}\label{OpL}
\mathcal{L}_i(g)= -\sigma_i \Big( g -M_i(v) \langle g \rangle \Big)= -\sigma_i\;g\quad \hbox{for} \; i=1,2    ,3.\end{equation} 
Using (\ref{equilibre}), (\ref{OpL}) and Lemma \ref{LE1}, then $\theta_i$ is given by
$$\theta_i= - \frac{1}{\sigma_i}v  M_i(v).$$
The other probability kernel $T_i^2$ is given by
\begin{equation*}
T_1^2[f_2](v,v^*)=\frac{\sigma_i\, D_1\,M_1\,v}{f_1}\big(1+\varphi_1(x,f_1)\big)\cdot\nabla\Big (\frac{f_1}{M_1}\Big)+K_{\frac{f_1}{M_1},\frac{f_2}{M_2}}(v,v^*)\cdot\nabla\Big (\frac{f_2}{M_2}\Big),
\end{equation*}

\begin{equation*}
T_2^2=\frac{\sigma_2\, D_2\,M_2\,v}{f_2}\big(1+\varphi_2(x,f_2)\big)\cdot\nabla\Big (\frac{f_2}{M_2}\Big),
\end{equation*}
and
\begin{equation*}
	T_3^2=\frac{\sigma_3\, D_3\,M_3\,v}{f_3}\big(1+\varphi_3(x,f_3)\big)\cdot\nabla\Big (\frac{f_3}{M_3}\Big),
\end{equation*}
where the functions $K_{\frac{f_2}{M_2},\frac{f_2}{M_2}}$ and $\varphi_i(x,f_i)$ satisfy the following asymptotic
$$K_{u_1+\varepsilon\frac{g_1}{M_1},u_2+\varepsilon\frac{g_2}{M_2}}=K_{u_1,u_2}+O(\varepsilon),\;\;\varepsilon\to 0,$$
$$\varphi_i(x,u_i+\varepsilon\frac{g_i}{M_i})=\varphi_i(x,u_i)+O(\varepsilon),\;\;\varepsilon\to 0.$$
From Eq. \eqref{3.4}, we obtain
\begin{equation*}
	\displaystyle\mathcal{T}_1^2[M_2u_2,M_3u_3](M_1)= -\frac{\sigma_1}{r^2}\,d\,|V|\chi(u_1,\,u_2) \cdot\,\, \nabla u_2,
\end{equation*}   
where $$\displaystyle\chi(u_1,u_2)=\Big\langle K_{u_1,u_2}(v,\,v^*)M_1(v)-K_{u_1,u_2}(v^*,\,v)M_1(v^*)\Big\rangle.$$
From \eqref{3.4} and Eq. \eqref{Gamma}, one has
$$\Gamma_1=\frac{D_1}{S}\big(1+\varphi_1(x,S)\big)\cdot\nabla S+\chi(S,I)\nabla I,$$
and
$$\Gamma_2=\frac{D_2}{I}\big(1+\varphi_2(x,I)\big)\cdot\nabla I,\qquad \Gamma_3=\frac{D_3}{R}\big(1+\varphi_2(x,R)\big)\cdot\nabla R.$$
\noindent Finally, the modeling of the interaction operators $G_{i}$ is given by 
\begin{equation}\label{GG1}
\left\{
\begin{array}{l}
\displaystyle
\displaystyle G_{1}(f_1,f_2,f_3)=  \frac{1}{|V|}\big(A-\mu f_1-\beta(t) f_1f_2/n\big),\\
\displaystyle G_{2}(f_1,f_2,f_3)=  \frac{1}{|V|}\big(\beta(t) f_1f_2/n-(\mu+\gamma)f_2\big),\\
\displaystyle G_{3}(f_1,f_2,f_3)=  \frac{1}{|V|}\big(\gamma f_2-\mu f_3\big).
\end{array}
\right.
\end{equation}
Then, using the definition of $H_i$ in (\ref{H}) to obtain from (\ref{GG1}) the following equality
\begin{equation} \label{HH}H_i(S,I,R)= F_i(S,I,R).\end{equation}
Collecting the previous results, we obtain the time-dependent nonlinear SIRD cross-diffusion system \eqref{Cross-Diffusion} of the order $O(\varepsilon)$

\begin{equation}\label{Cross-Diffusion2}
\begin{cases}
\displaystyle \partial_t S=d_1\nabla \cdot(\varphi_1(x,S) \nabla S)+ \nabla\bigl(\chi(S,I)\Grad
I\bigr)+A-\mu S-\beta(t)SI/N+O(\varepsilon),\\
{}\\
\displaystyle \partial_t I=d_2\nabla \cdot(\varphi_2(x,I) \nabla I)+\beta(t)SI/N-(\mu+\gamma)I+O(\varepsilon),
\\
{}\\
\displaystyle \partial_t R=d_3\nabla \cdot(\varphi_3(x,R) \nabla R)+\gamma I-\mu R+O(\varepsilon),\\
{}\\
\displaystyle \partial_t D=\alpha I.
\end{cases}
\end{equation}
\section{Numerical analysis of the equivalent micro-macro formulation in one dimensional space}\label{Sec4}
In this section, we develop an asymptotic preserving (AP)-scheme in one dimension of the equivalent micro-macro formulation developed in Section \ref{Sec3}. This method designs uniform stability with respect to the parameter $\varepsilon$, related to the mean distance between individuals, as well as consistency with the nonlinear cross-diffusion limit. The discretization of micro-macro formulation (\ref{mM1}) is carried out with respect to each independent variable, namely time, space and  velocity.
\subsection{Semi-implicit time discretization}
Here we present a time discretization of micro-macro formulation (\ref{mM1}). Let denote by $\Delta t$ a fixed time step, and by $t_k$  a discrete time such that $t_k=k\, \Delta t $  $ k\in N.$ The approximation of $u_i(t,x)$ and $g_i(t,x,v)$ at the time step $t_k$ are denoted respectively by $u_i^k\approx u_i(t_k,x)$ and $g_i^k\approx g_i(t_k,x,v)$.\\
 In the first microscopic equations of (\ref{mM1}), the term  $\displaystyle  \frac{1}{\varepsilon}\mathcal{L}_i(g_i)$ presents a stiffness in the collision part for small $\varepsilon$. Thus, it is natural to take an  implicit scheme to ensure the stability for this term, while  the other terms are still  explicit. Consequently,
\begin{equation}\label{mMD1}
\begin{array}{ll}
\displaystyle  \frac{ g_i^{k+1}-g_i^{k}}{\Delta t} +
\frac{1}{\varepsilon^2} v M_i \cdot \nabla u_i^k + \frac{1}{\varepsilon} (I-P_{M_i})(v\cdot\nabla g_i^k)=  \frac{1}{\varepsilon^{2}}\mathcal{L}_i(g_i^{k+1})  \\\\
\hskip1cm\displaystyle+\frac{1}{\varepsilon^2}\mathcal{T}_i^2[M_1u_1^k,\dots,M_{i-1}u_{i-1}^k, M_{i+1}u_{i+1}^k,\dots,M_3u_3^k](M_iu_i^k)\\\\
\displaystyle \hskip1cm+\frac{1}{\varepsilon}\mathcal{T}_i^2[M_1u_1^k,\dots,M_{i-1}u_{i-1}^k, M_{i+1}u_{i+1}^k,\dots,M_3u_3^k](g_i^k)\\ \\ 
\displaystyle \hskip1cm
+\frac{1}{\varepsilon} (I-P_{M_i})G_i(u_1^k,u_2^k,u_3^k).
\end{array}
\end{equation}
In the second  macroscopic equations of (\ref{mM1}), we take the function $g$ at the time $t_{k+1}$, which gives
\begin{equation}\label{mMD2}
\frac{ u_i^{k+1}-u_i^{k}}{\Delta t}  + \langle v    \cdot
\nabla g_i^{k+1} \rangle = \left\langle G_i(u_1^k ,u_2^k,u_3^k)\right\rangle.
\end{equation}

\begin{proposition}
	The time discretization (\ref{mMD1})-(\ref{mMD2}) is consistent with (\ref{G1}) in the limit.
\end{proposition}
\subsection{Fully discrete asymptotic preserving (AP)-scheme in 1D}
Here we construct a suitable space discretization of (\ref{mMD1})-(\ref{mMD2}) using finite volume method. The domain space under consideration is $[-L,L]$. Note that the velocity space in the interval $[-V,V]$ can be treated by using a standard discretization. \\

For this, let denote by $K_j=]x_{j-\frac{1}{2}},x_{j+\frac{1}{2}}[$ the control volume where $x_j=\frac{1}{2}(x_{j-\frac{1}{2}}+x_{j+\frac{1}{2}})$ and its length is denoted by  $h_j=x_{j+\frac{1}{2}}-x_{j-\frac{1}{2}}$ for $ j=1,\dots    ,N_x$, ($N_x$ is the total number of cells).  The approach consists to compute the macroscopic densities in $K_j$ and the microscopic quantities are computed on $\partial K_j$ as follow
$$u_i(t_k,x)\arrowvert_{K_j}\approx u_{i,j}^k, \;\; \hbox{and}\;\;  g_i(t_k,x_{j+\frac{1}{2}},v)\arrowvert_{\partial K_j}\approx g_{i, j+\frac{1}{2}}^k,\;\; i=1,\dots    ,3,\; j=1,\dots    ,N_x.$$
Then, the full discretization of the equivalent micro-macro formulation \eqref{mM1} is given as follow
\begin{equation}\label{mMDD1}
\begin{array}{ll}
\displaystyle  \frac{ g^{k+1}_{i,j+\frac{1}{2}}-g^{k}_{i,j+\frac{1}{2}}}{\Delta t} +
\frac{1}{\varepsilon^2}  v M\frac{u^k_{i,j+1} -u^k_{i,j}}{h_j} + \frac{1}{\varepsilon} (I-P_{M_i})\Big(v^+\frac{g^k_{i,j+\frac{1}{2}}-g^k_{i,j-\frac{1}{2}}}{h_j} +v^-\frac{g^k_{i,j+\frac{3}{2}}-g^k_{i,j+\frac{1}{2}}}{h_j} \Big) \\\\
\hskip0.cm\displaystyle =  \frac{1}{\varepsilon^{2}}\mathcal{L}_i(g^{k+1}_{i,j+\frac{1}{2}})
+\frac{1}{\varepsilon^2}\mathcal{T}_i^2[M_1u_{1,j+\frac{1}{2}}^k,\dots,M_{i-1}u_{i-1,j+\frac{1}{2}}^k, M_{i+1}u_{i+1,j+\frac{1}{2}}^k,\dots,M_3u_{3,j+\frac{1}{2}}^k](M_iu_{i,j+\frac{1}{2}}^k)\\\\
\displaystyle \hskip1cm+\frac{1}{\varepsilon}\mathcal{T}_i^2[M_1u_{1,j+\frac{1}{2}}^k,\dots,M_{i-1}u_{i-1,j+\frac{1}{2}}^k, M_{i+1}u_{i+1,j+\frac{1}{2}}^k,\dots,M_3u_{3,j+\frac{1}{2}}^k](g_{i,j+\frac{1}{2}}^k) \\\\
\displaystyle \hskip1cm +\frac{1}{\varepsilon}(I-P_{M_i})G_i(u^k_{1,j+\frac{1}{2}},\cdots,u^k_{3,j+\frac{1}{2}}),\\\\
\displaystyle
\frac{ u^{k+1}_{i,j}-u^{k}_{i,j}}{\Delta t}  + \Big\langle v  \frac{ g^{k+1}_{i,j+\frac{1}{2}}-g^{k+1}_{i,j-\frac{1}{2}}}{h_j} \Big\rangle = \langle G_i(u^k_{1,j},\cdots,u^k_{3,j})
\rangle,
\end{array}
\end{equation}
where $u_{i,j+\frac{1}{2}}=\frac{u_{i,j+1}+u_{i,j}}{2}$ and $u_{i,j-\frac{1}{2}}=\frac{u_{i,j}+u_{i,j-1}}{2}$.

\begin{proposition}\label{Pro3.2}
	The time and space approximation  (\ref{mMDD1}) of kinetic equation (\ref{so}) in the limit $\varepsilon$ goes to zero satisfy the following discretization
	\begin{equation}\label{mMD221}
	\begin{array}{ll}
	\displaystyle
	\frac{ u_i^{k+1}-u_i^{k}}{\Delta t}  + \frac{1}{h_j}\Big\langle  v\cdot  \Big [ \mathcal{L}_i^{-1}\Big( v M(v)  \frac{u^k_{i,j+1} -u^k_{i,j} }{h_j} + v M(v)  \frac{u^k_{i,j} -u^k_{i,j-1} }{h_j} \\\\
	\displaystyle\hskip0.2cm-\mathcal{T}_i^2[M_1u_{1,j+\frac{1}{2}}^k,\dots,M_{i-1}u_{i-1,j+\frac{1}{2}}^k, M_{i+1}u_{i+1,j+\frac{1}{2}}^k,\dots,M_3u_{3,j+\frac{1}{2}}^k](M_iu_{i,j+\frac{1}{2}}^k) 
	\\\\
	\displaystyle\hskip0.2cm-\mathcal{T}_i^2[M_1u_{1,j-\frac{1}{2}}^k,\dots,M_{i-1}u_{i-1,j-\frac{1}{2}}^k, M_{i+1}u_{i+1,j-\frac{1}{2}}^k,\dots,M_3u_{3,j-\frac{1}{2}}^k](M_iu_{i,j-\frac{1}{2}}^k)\Big)\Big ] \Big\rangle \\\\\displaystyle\hskip0.2cm =\langle  G_i(u^k_{1,j+\frac{1}{2}},\cdots,u^k_{3,j+\frac{1}{2}})\rangle,
	\end{array}
	\end{equation}
	which is consistent with the first equation of (\ref{x1}).
\end{proposition}

\subsection{Boundary conditions}{\label{SUBBC}}
For the numerical solution of the kinetic equation (\ref{Cross-Diffusion}),  usually the inflow boundary conditions are prescribed as follows 
$$f_i(t,x_{\min},v)=f_{i,l}(v),\quad v>0,\qquad f_i(t,x_{\max},v)=f_{i,r}(v),\quad v<0, \quad \hbox{for} \; i=1,...,3.$$
Thus, the inflow boundary conditions can be rewritten in the micro-macro formulation \eqref{mM1} as follow
$$\displaystyle u_i(t,x_0)M_i(v)+\frac{\varepsilon}{2}(g_i(t,x_{\frac{1}{2}},v)+g_i(t,x_{-\frac{1}{2}},v))=f_{i,l}(v),\quad v<0,$$
$$\displaystyle u_i(t,x_{N_x})M_i(v)+\frac{\varepsilon}{2}(g_i(t,x_{N_x+\frac{1}{2}},v)+g_i(t,x_{N_x-\frac{1}{2}},v))=f_{i,r}(v),\quad v>0.$$
We consider the following artificial Neumann boundary conditions for the other velocities
$$\displaystyle g_i(t,x_{\frac{1}{2}},v)=g_i(t,x_{-\frac{1}{2}},v),\quad v<0,$$
$$\displaystyle g_i(t,x_{N_x+\frac{1}{2}},v)=g_i(t,x_{N_x-\frac{1}{2}},v),\quad v>0.$$
Furthermore, the ghost points can be computed as follows
\begin{equation}
\displaystyle g_{i,j-\frac{1}{2}}^{k+1}=\left\{
\begin{array}{ll}
\displaystyle\frac{2}{\varepsilon}\Big(f_{i,l}(v)-u_{i,0}^{k+1}M_i(v)\Big)-g_{i,\frac{1}{2}}^{k+1},\quad& v>0,  \\
{}  \\
\displaystyle g_{i,\frac{1}{2}}^{k+1},\quad& v<0,
\end{array} \right.
\end{equation}
\begin{equation}
\displaystyle g_{i,N_x+\frac{1}{2}}^{k+1}=\left\{
\begin{array}{ll}
\displaystyle\frac{2}{\varepsilon}\Big(f_{i,r}(v)-u_{i,N_x}^{k+1}M_i(v)\Big)-g_{i,N_x-\frac{1}{2}}^{k+1},\quad &v<0,  \\
{}  \\
\displaystyle g_{i,N_x-\frac{1}{2}}^{k+1},\quad& v>0.
\end{array} \right. 
\end{equation}
Finally, we use (\ref{mMDD1}) to obtain 
\begin{equation}
\left\{
\begin{array}{l}
\displaystyle \Big(1+\frac{2\Delta t}{\varepsilon \Delta x}\langle v^+M_{i}(v)\rangle\Big)u_{i,0}^{k+1}=u_{i,0}^k-\frac{\Delta t}{\Delta x} \Big\langle(v+v^+-v^-)g_{i,\frac{1}{2}}^{k+1}-\frac{2v_l^+}{\varepsilon}f_{l}(v)\Big\rangle\\\\
\displaystyle\hskip5cm+\Delta t \,G_i(u_{1,0}^k,\cdots,u_{3,0}^k),\\
\\
{}  \\
\displaystyle \Big(1-\frac{2\Delta t}{\varepsilon \Delta x}\langle v^-M_{i}(v)\rangle\Big)u_{i,N_x}^{k+1}=u_{i,N_x}^k-\frac{\Delta t}{\Delta x} \big\langle\frac{2v^-}{\varepsilon}f_{r}(v)-(v-v^++v^-)g_{i,N_x-\frac{1}{2}}^{k+1}\big\rangle\\\\
\displaystyle \hskip5cm+\Delta t\,G_i(u_{1,N_x}^k,\cdots,u_{3,N_x}^k). 
\end{array} \right. 
\end{equation}

\subsection{Numerical simulations}
We provide some numerical simulations obtained with the equivalent micro-macro formulation presented in Section \ref{Sec4} and from the macroscopic scheme. First, we show the asymptotic preservation scheme property. Second, we provide the role of the transmission function $ \beta(t) $. Next, we demonstrate the effect of the diffusion terms on the evolution of the individuals. Finally, we show the role of the presence of the cross-diffusion term by different choices of the function $\chi(S,I)$.

We consider that the velocity space is the interval $ V = [-1.1] $ with the number of grids $ N_v = 164 $, which can provide sufficient precision for numerical simulations \cite{[KY13]}. The step time is $ t = 10^ {- 3} $ and the space domain is the interval $ \Omega = [-2, 2] $ with the number of cells $ N_x = 200 $. We take the following set of parameters as an example to analyze the results by varying some of them: $ \mu = 1/83, \;  \; \gamma = 1/3,\; \chi=0.01$. Three cases of the diffusion coefficients are considered: i) without diffusion ($ d_i = 0 $), ii) with diffusion, namely $ d_1 = 0.05, \; d_2 = 0.025, \; d_3 = $ 0.001 and $ d_4 = $ 0, the same as Reference \cite {[SSL10]} where the functions $ \varphi_i (x) = \mid x\mid$, and iii) same as the case ii) with diffusion coefficient depending on $x$, namely $  \varphi_i (x) = 1+0.5\,x$. Finally, we take the following initial conditions:
\begin{itemize}
	\item[$i)$]
	\begin{equation*}
	\left\{
	\begin{array}{l}
	S_0=2.6\,\Big(\exp(-(\frac{x-0.5}{0.12})^2)+\exp(-(\frac{x+0.5}{0.12})^2)\Big)/(0.9\,\pi),\\
	I_0=0.04\,\exp(-2\,x^2),	\\
	R_0=0, \\
	N_0=S_0+I_0, 
	\end{array} \right. 
	\end{equation*}
	
	\item[$ii)$] 	
	\begin{equation*}
	\left\{
	\begin{array}{l}
	S_0=0.96\,\exp(-10(\frac{x}{1.4})^2),\\
	I_0=0.04\,\exp(-2\,x^2),	\\
	R_0=0, \\
	N_0=S_0+I_0. 
	\end{array} \right. 
	\end{equation*}
\end{itemize}

\subsubsection{Test 1: Asymptotic preserving property}
In this test we aim to validate the asymptotic preserving numerical scheme property. We consider the initial conditions $ i) $, the diffusion case  $b)$  and the reproduction ratio is $ R_0 = 2$.

\begin{figure}[h!]

	\subfigure[]{ \includegraphics[height=1.8in ,width=2in]{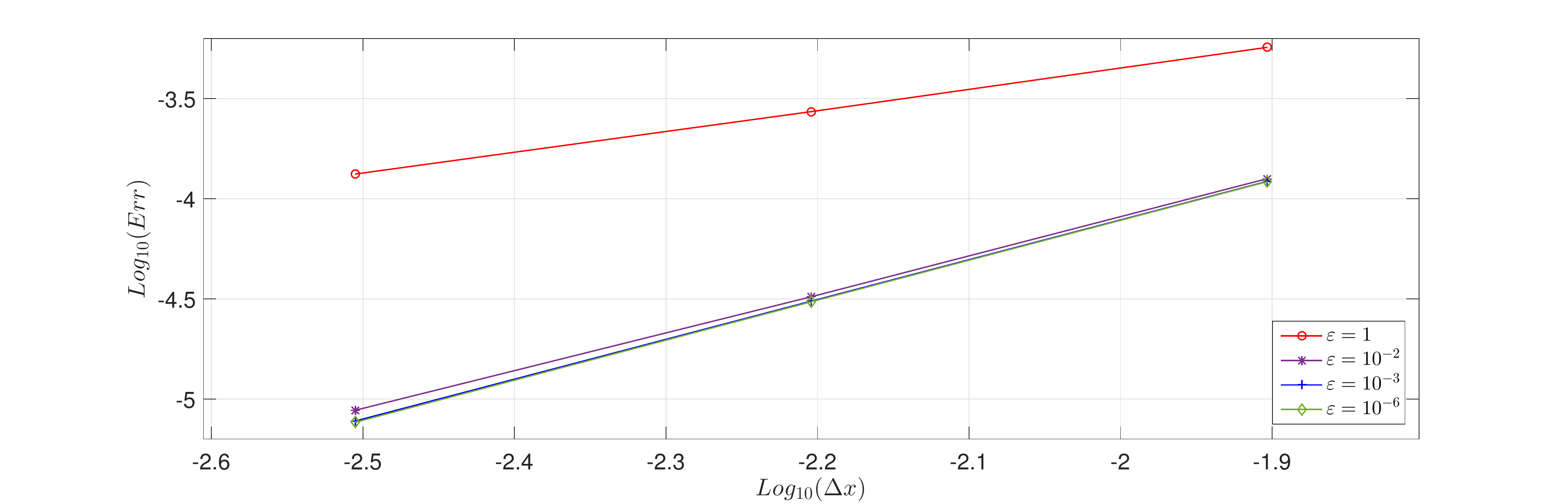}}
	~\subfigure[]{\includegraphics[height=1.8in ,width=2in]{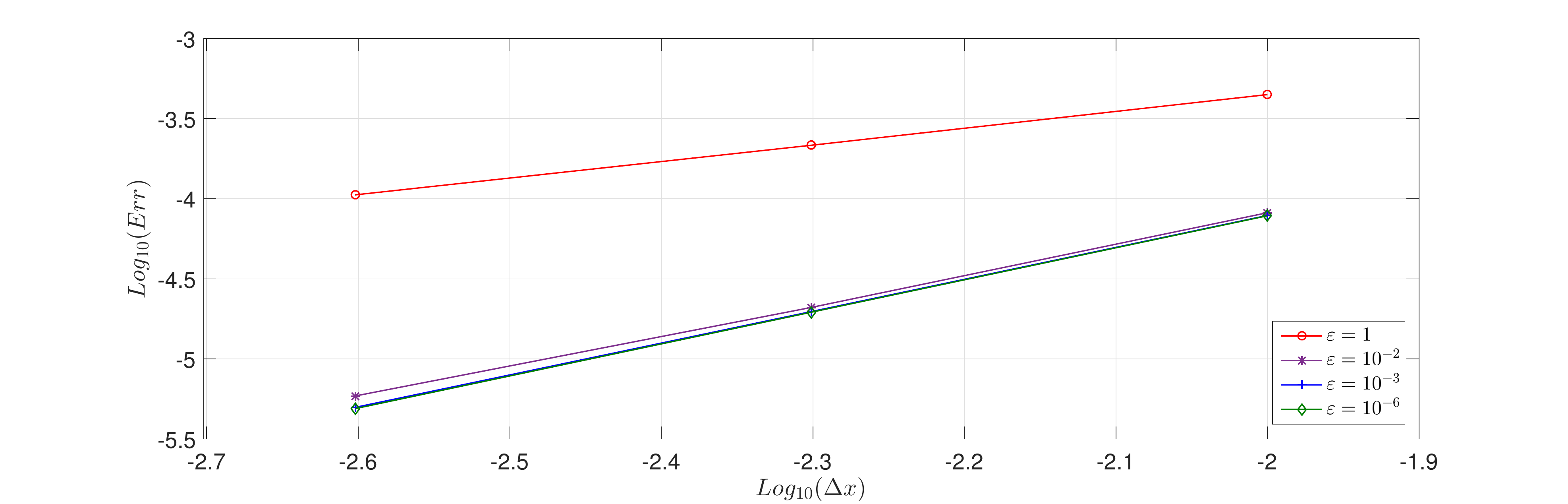}}
	~\subfigure[]{\includegraphics[height=1.8in ,width=2in]{error2.pdf}}
	\caption{Convergence order of the method for $\varepsilon=\{1, 10^{-2}, 10^{-3}, 10^{-6}\}$ at time $t = 0.01$ ($M = 1$) for the density $S$ in the left, the density $I$ in the middle and the density $R$ in the right obtained from the asymptotic preserving numerical scheme.}
	\label{Err}
\end{figure}

 In Figure \ref{Err}, we present the plots in log scale of the error estimates given by 
$$ e_{\Delta x} (h)= \frac{|h_{\Delta x} (t)- h_{2\Delta x} (t)|_1}{| h_{2\Delta x} (0)|_1} $$
to test the convergence of our scheme. This can be considered as an estimation of the relative error in $l^1$ norm, where $h_{\Delta x}$ is the numerical solution computed from a uniform grid of size $\displaystyle\Delta x= \frac{x_{max}-x_{min}}{N_x}$. The computations are performed with $N_x=\{80, 160, 320, 640 \}  ,$ $\displaystyle\Delta t= 10^{-6} $ at $t=0.01$ for $\displaystyle\varepsilon =\{ 1,10^{-2},10^{-3},10^{-6} \}$. 

\begin{figure}
	\centering
	\begin{tikzpicture}[spy using outlines={ ,blue,magnification=3,size=1.5cm, connect spies}]
		\node{\includegraphics[height=1.7in ,width=1.8in]{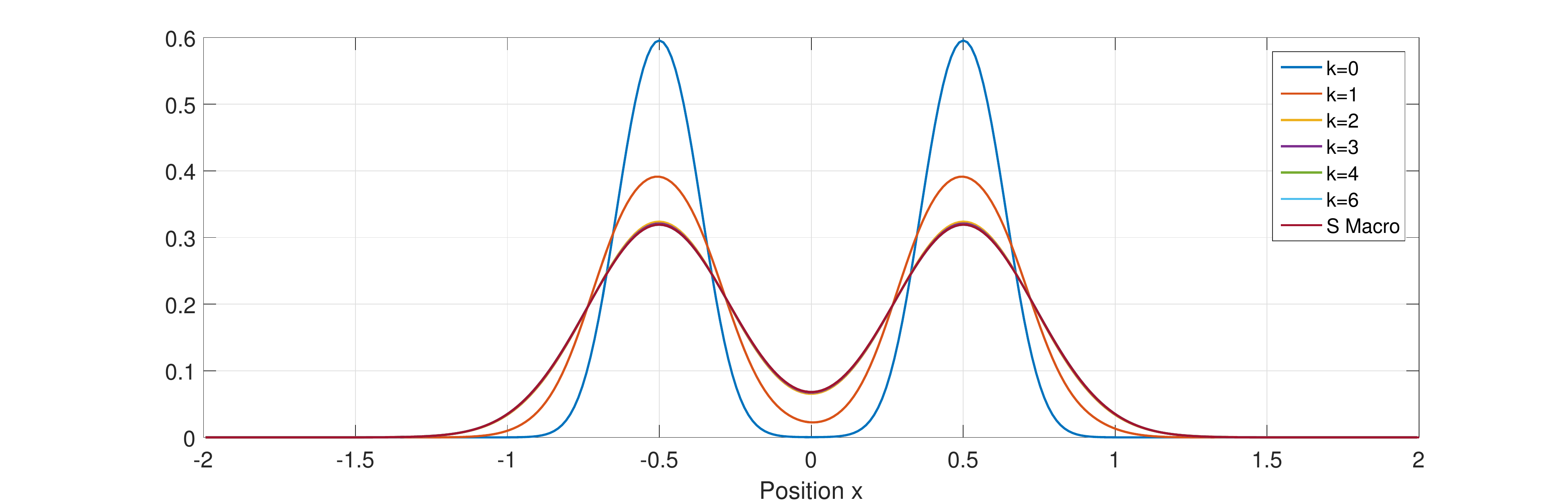}};
		\spy on (0.1,-1.2) in node [right] at  (.5,-1.45);
	\end{tikzpicture}
	\begin{tikzpicture}[spy using outlines={ ,blue,magnification=3,size=1.5cm, connect spies}]
		\node{\includegraphics[height=1.7in ,width=1.8in]{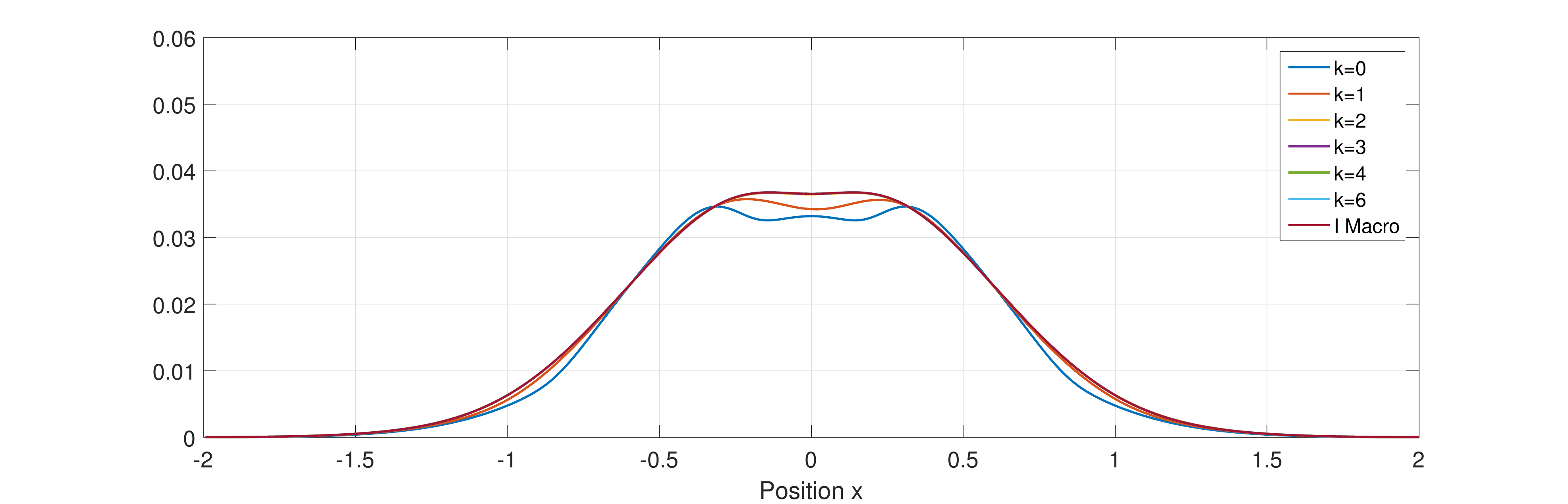}};
		\spy on (0.1,0.4) in node [right] at  (.5,-1.45);
	\end{tikzpicture}
	\begin{tikzpicture}[spy using outlines={ ,blue,magnification=3,size=1.5cm, connect spies}]
		\node{\includegraphics[height=1.7in ,width=1.8in]{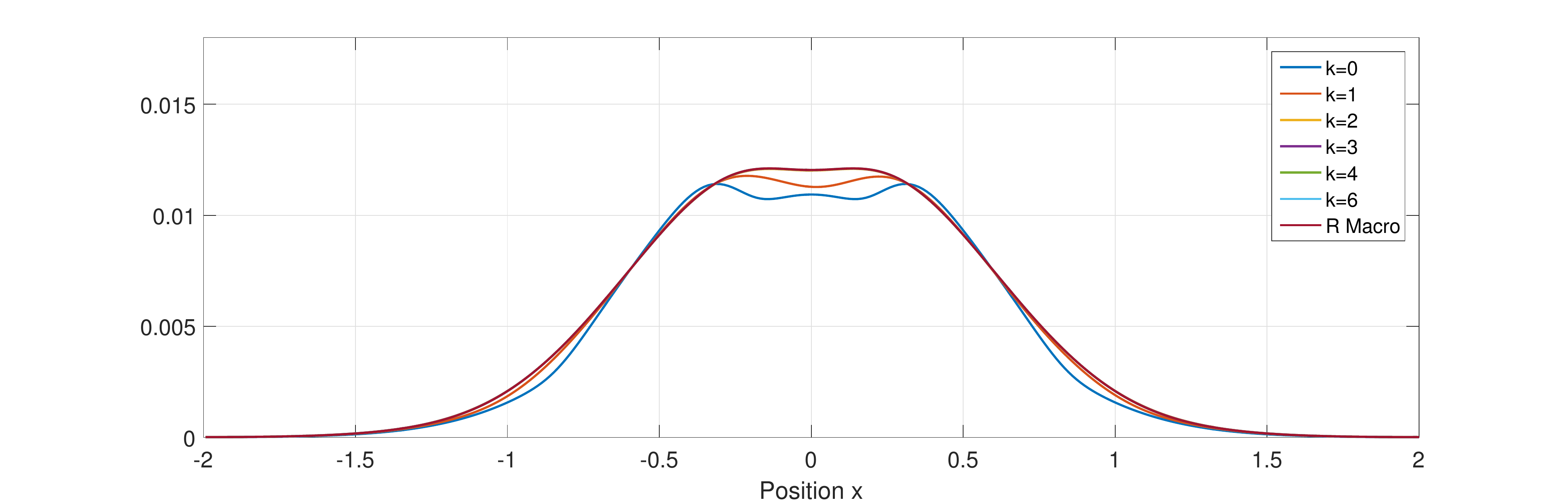}};
		\spy on (0.1,0.55) in node [right] at  (.5,-1.45);
	\end{tikzpicture}	
	\begin{tikzpicture}[spy using outlines={ ,blue,magnification=3,size=1.5cm, connect spies}]
	\node{\includegraphics[height=1.7in ,width=1.8in]{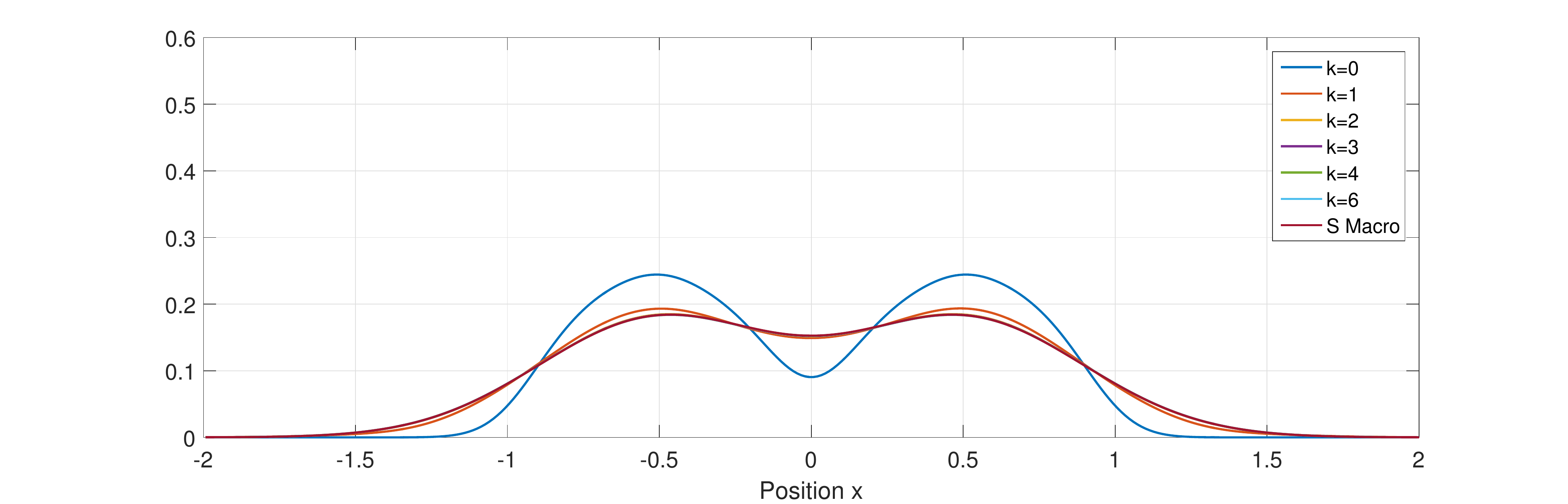}};
	\spy on (0.1,-0.7) in node [right] at  (.5,-1.45);
\end{tikzpicture}
\begin{tikzpicture}[spy using outlines={ ,blue,magnification=3,size=1.5cm, connect spies}]
	\node{\includegraphics[height=1.7in ,width=1.8in]{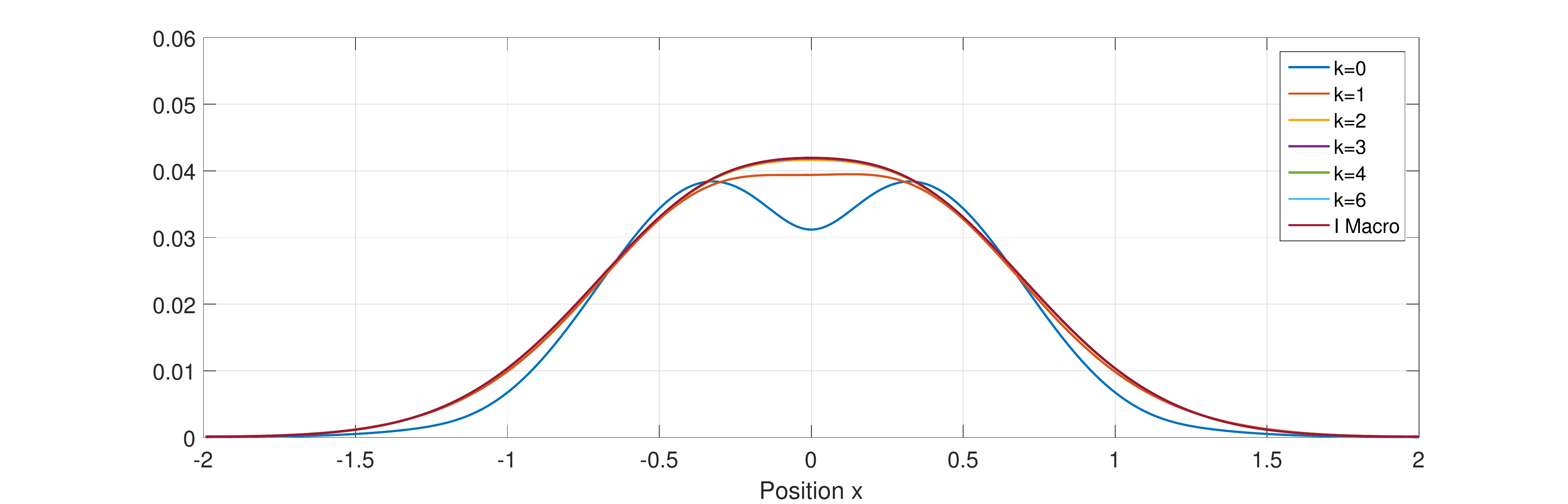}};
	\spy on (0.1,0.65) in node [right] at  (.5,-1.45);
\end{tikzpicture}
\begin{tikzpicture}[spy using outlines={ ,blue,magnification=3,size=1.5cm, connect spies}]
	\node{\includegraphics[height=1.7in ,width=1.8in]{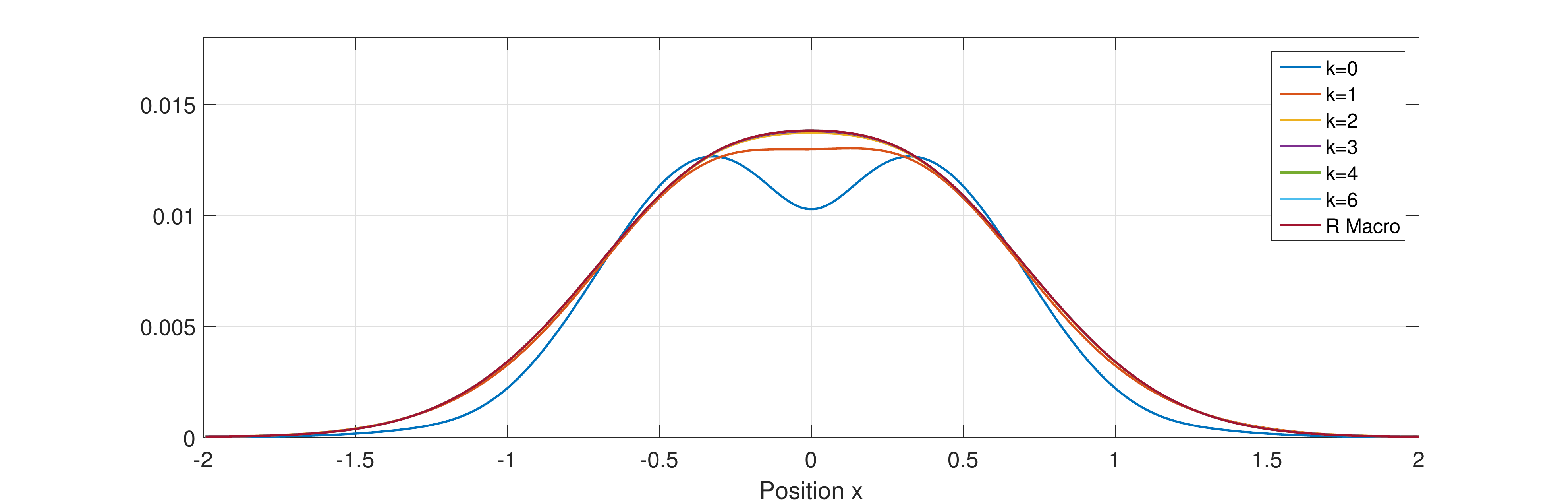}};
	\spy on (0.1,0.85) in node [right] at  (.5,-1.45);
\end{tikzpicture}
	\begin{tikzpicture}[spy using outlines={ ,blue,magnification=3,size=1.5cm, connect spies}]
	\node{\includegraphics[height=1.7in ,width=1.8in]{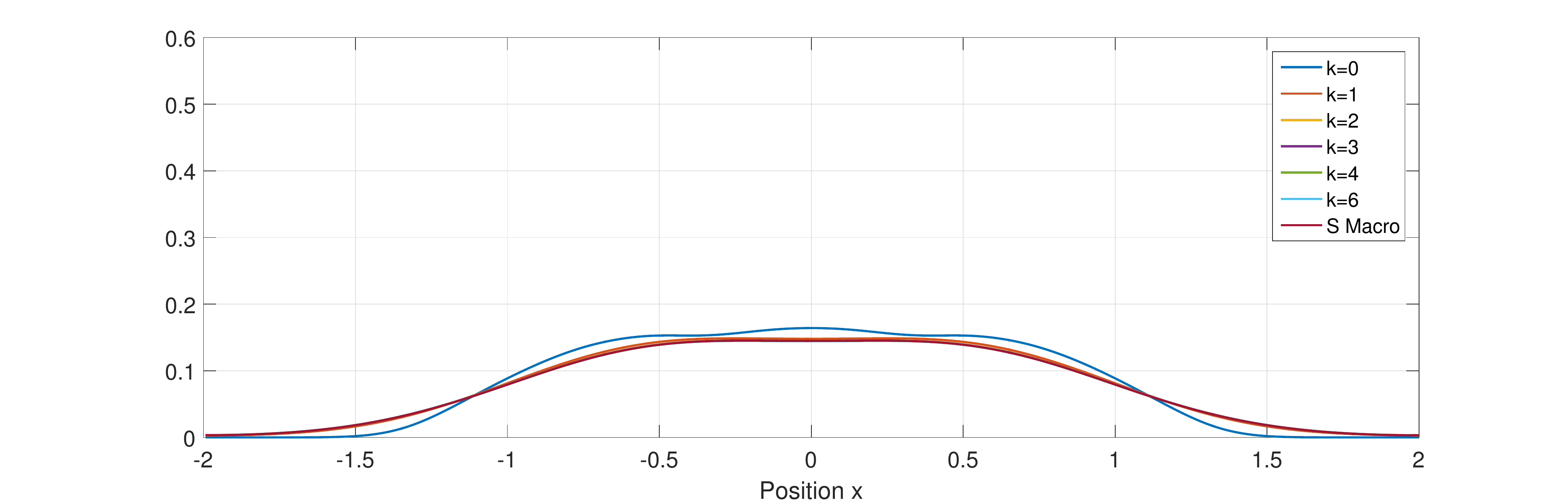}};
	\spy on (0.1,-0.7) in node [right] at  (.5,-1.45);
\end{tikzpicture}
\begin{tikzpicture}[spy using outlines={ ,blue,magnification=3,size=1.5cm, connect spies}]
	\node{\includegraphics[height=1.7in ,width=1.8in]{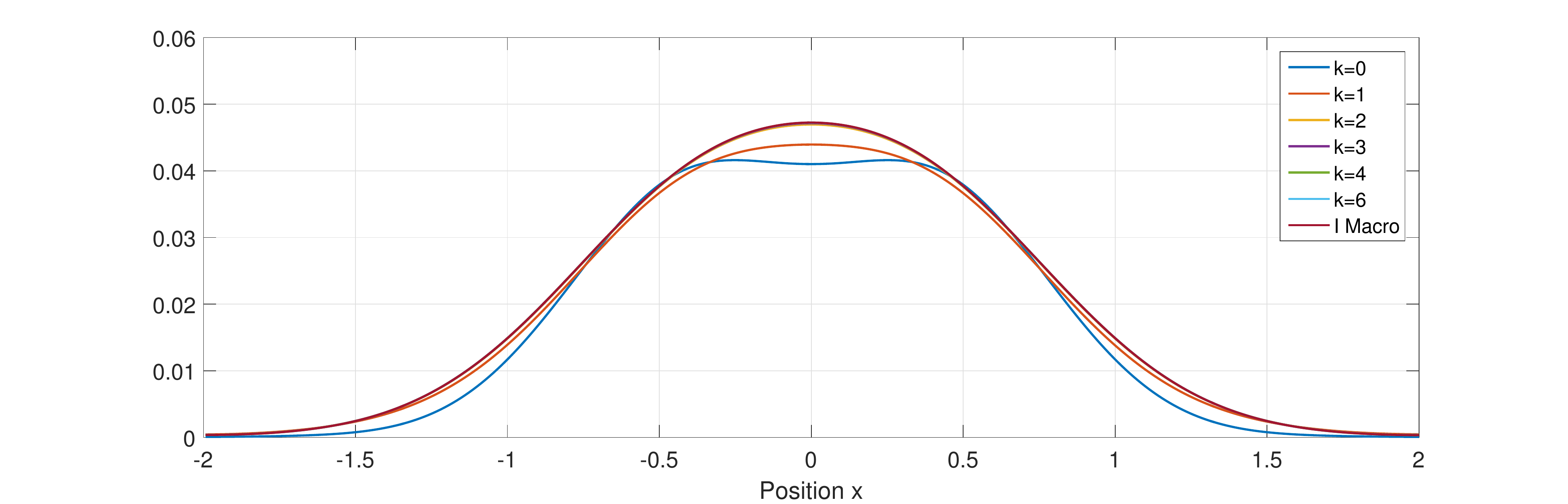}};
	\spy on (0.1,0.9) in node [right] at  (.5,-1.45);
\end{tikzpicture}
\begin{tikzpicture}[spy using outlines={ ,blue,magnification=3,size=1.5cm, connect spies}]
	\node{\includegraphics[height=1.7in ,width=1.8in]{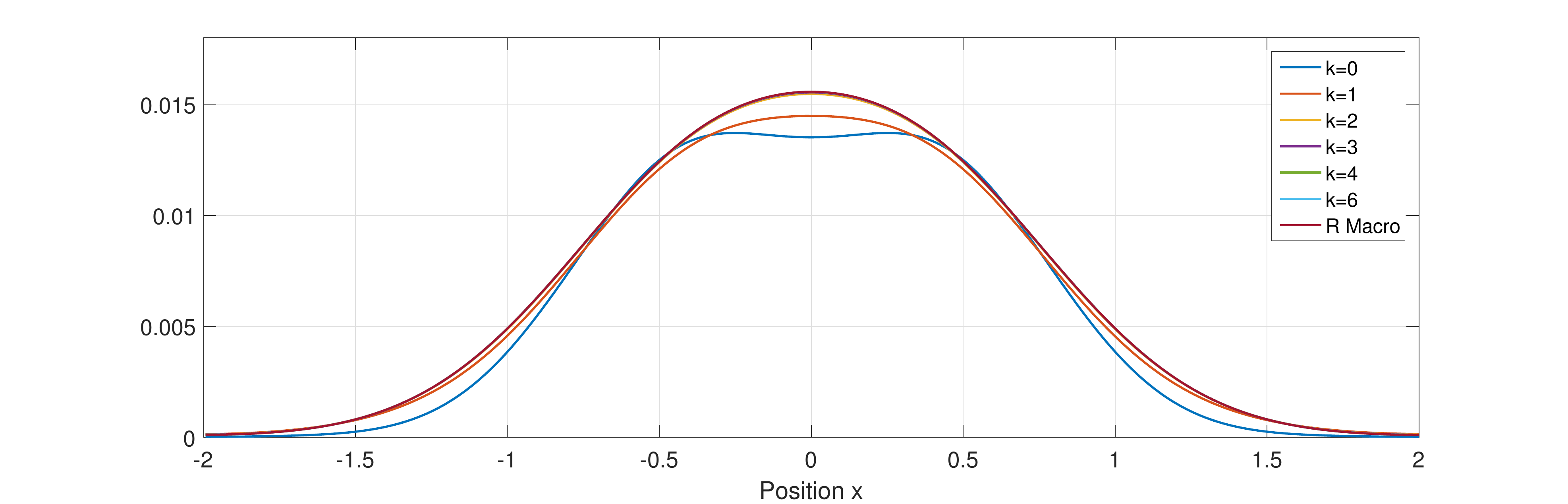}};
	\spy on (0.1,1.2) in node [right] at  (.5,-1.45);
\end{tikzpicture}
	\begin{tikzpicture}[spy using outlines={ ,blue,magnification=3,size=1.5cm, connect spies}]
	\node{\includegraphics[height=1.7in ,width=1.8in]{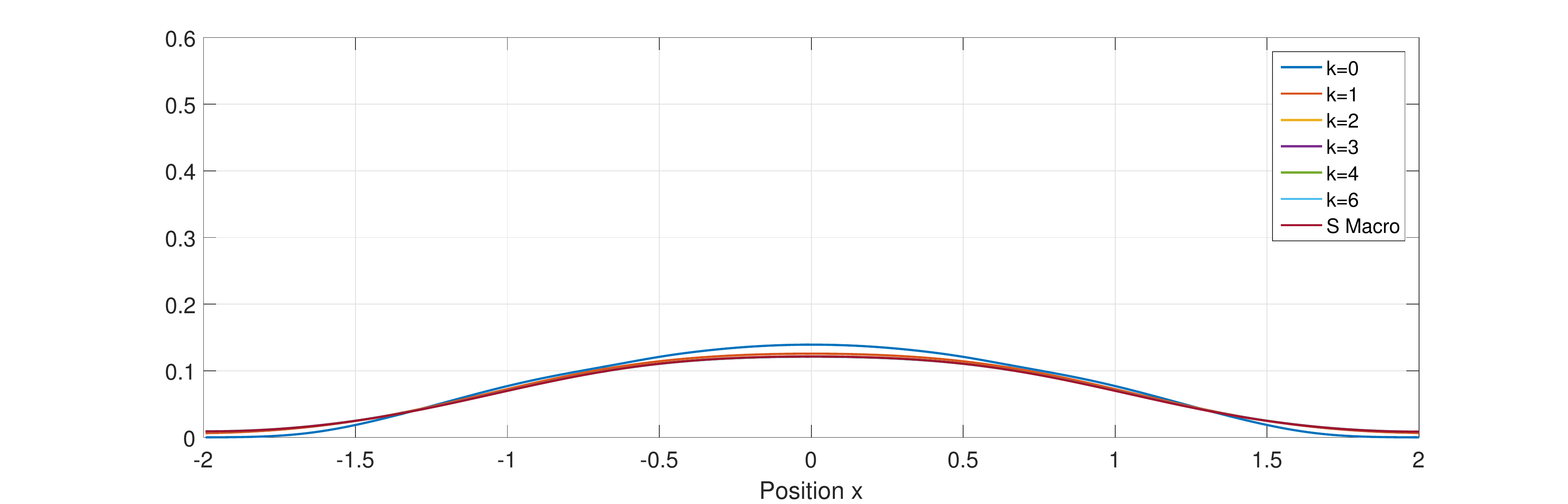}};
	\spy on (0.1,-0.8) in node [right] at  (.5,-1.45);
\end{tikzpicture}
\begin{tikzpicture}[spy using outlines={ ,blue,magnification=3,size=1.5cm, connect spies}]
	\node{\includegraphics[height=1.7in ,width=1.8in]{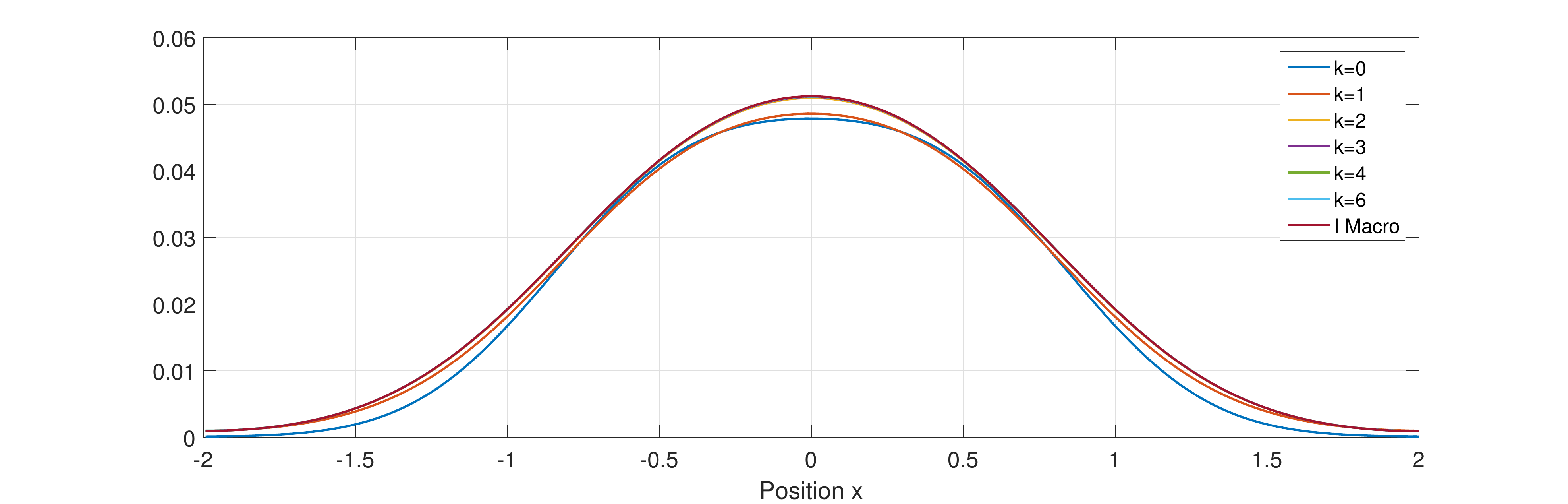}};
	\spy on (0.1,1.2) in node [right] at  (.5,-1.45);
\end{tikzpicture}
\begin{tikzpicture}[spy using outlines={ ,blue,magnification=3,size=1.5cm, connect spies}]
	\node{\includegraphics[height=1.7in ,width=1.8in]{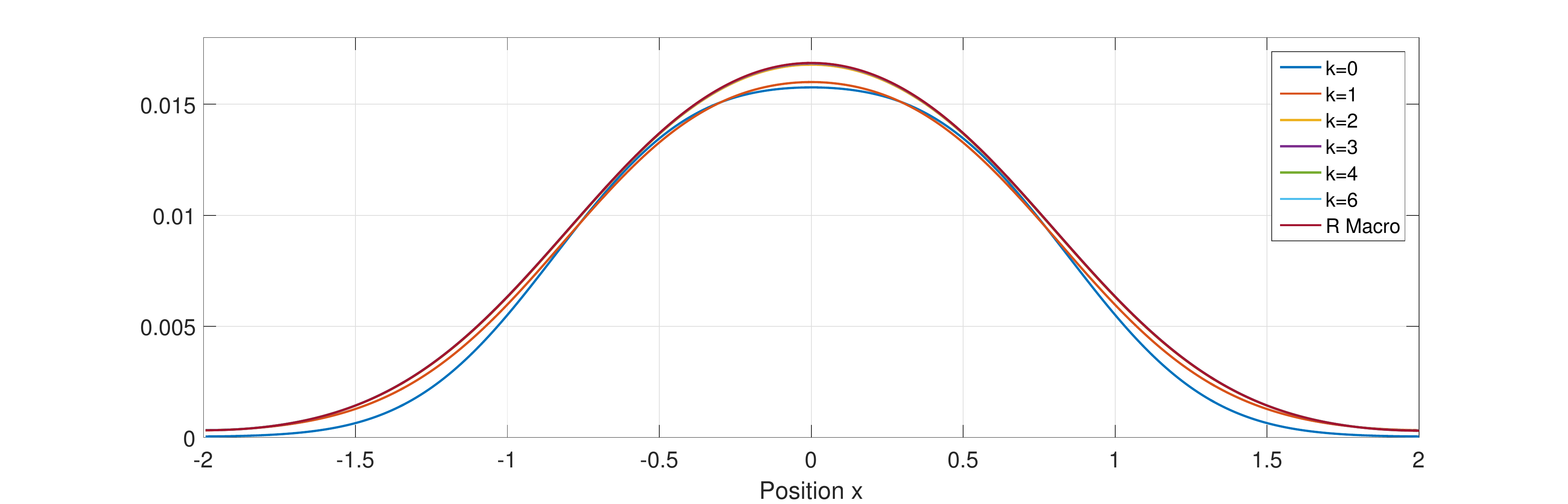}};
	\spy on (0.1,1.5) in node [right] at  (.5,-1.45);
\end{tikzpicture}
	\caption{Dynamics of the densities $f_1$ (first column), $f_2$ (second column) and $f_3$ (third column) obtained with the asymptotic preserving numerical scheme for $\varepsilon =2\times10^{-k}$, $k = 0,\,1,\, 2,\,3,\, 4,\, 6$ and with the macroscopic numerical scheme using initial conditions $i)$ at successive time $t=0.5,\,1,\,5,\,10$.}\label{F0}	
\end{figure}

Figure \ref{F0} shows the numerical results  of susceptible, infected and recovered individuals obtained with micro-macro scheme presented in Sec. \ref{Sec4}  and with macroscopic numerical scheme  at successive instants $ t = 0.5, \, 1, \, 5, \, $ 10. The obtained numerical results have almost the same profiles in the limit when the parameter $ \displaystyle \varepsilon = 2 \times10 ^ {- k} $, with $ k = 0, \, 1, \, 2, \, 3, \, 4, \, 6 $, goes to zero. This confirms that the asymptotic preserving numerical scheme is uniformly stable along the transition from kinetic regime to macroscopic regime, which illustrates the result in Proposition \ref{Pro3.2}.

\subsubsection{Test 2: Time-dependent effect by $\beta(t)$}
\begin{figure}[h!]
	\centering
	\subfigure[]{\includegraphics[height=1.7in ,width=2in]{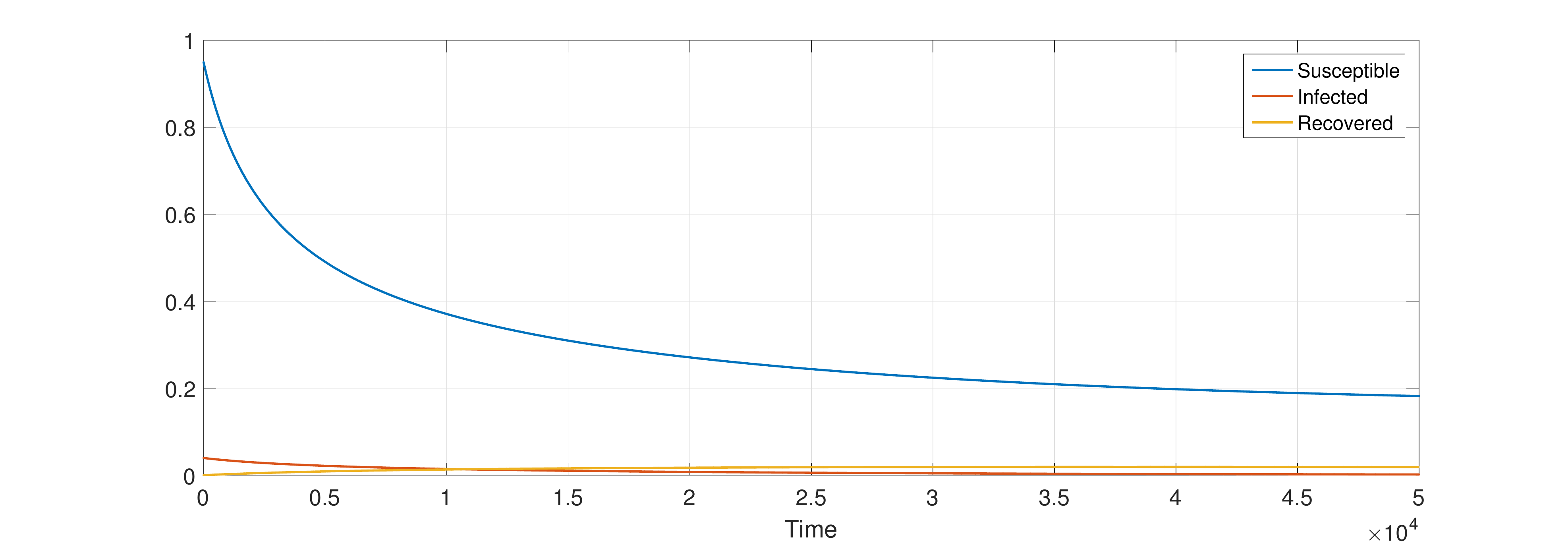}}~
	\subfigure[]{\includegraphics[height=1.7in ,width=2in]{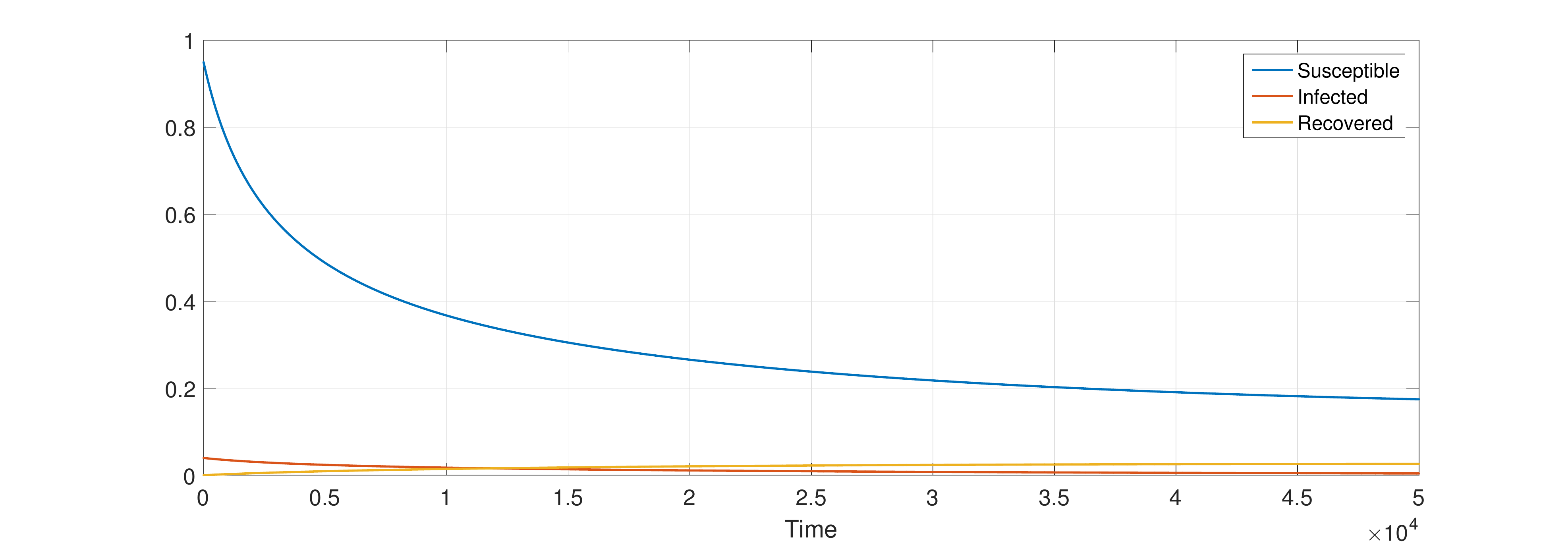}}~
	\subfigure[]{\includegraphics[height=1.7in ,width=2in]{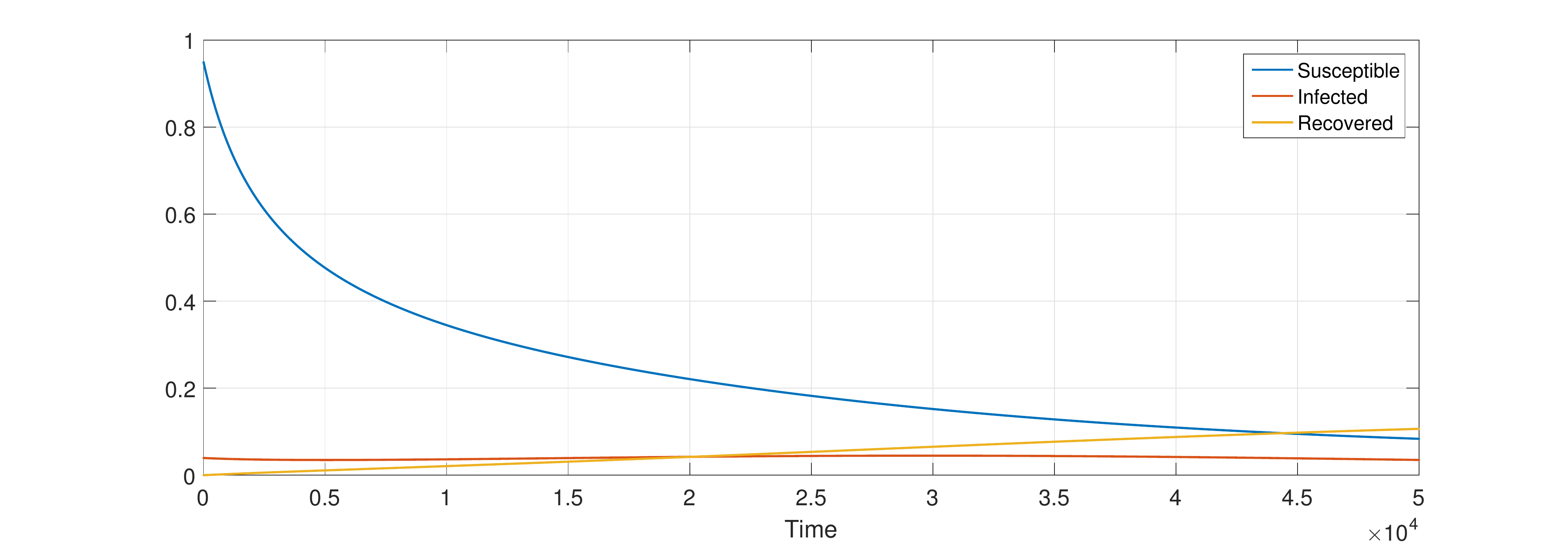}}
	\subfigure[]{\includegraphics[height=1.7in ,width=2in]{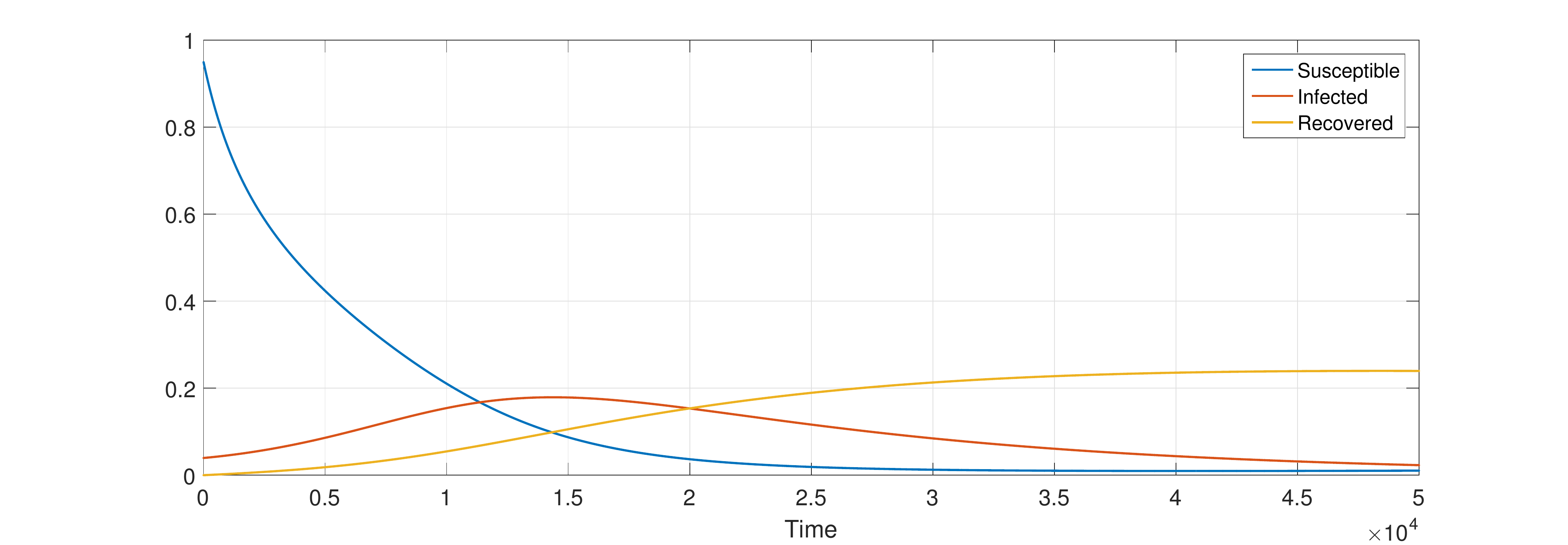}}~
	\subfigure[]{\includegraphics[height=1.7in ,width=2in]{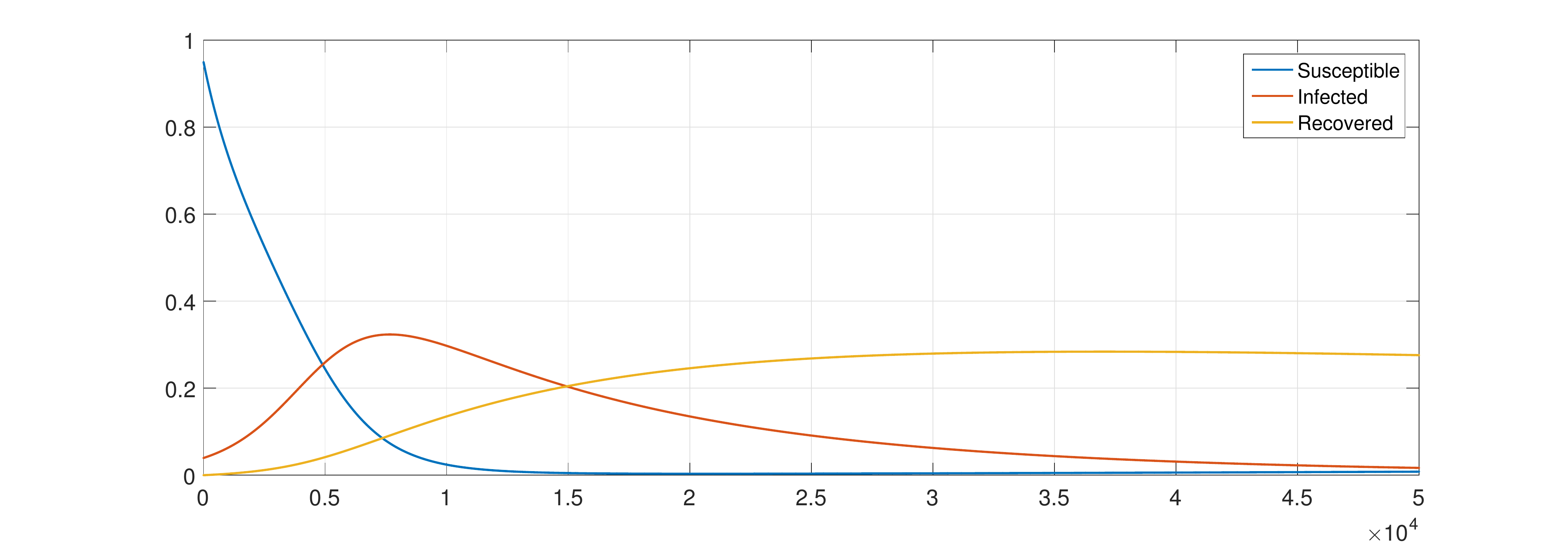}}~
	\subfigure[]{\includegraphics[height=1.7in ,width=2in]{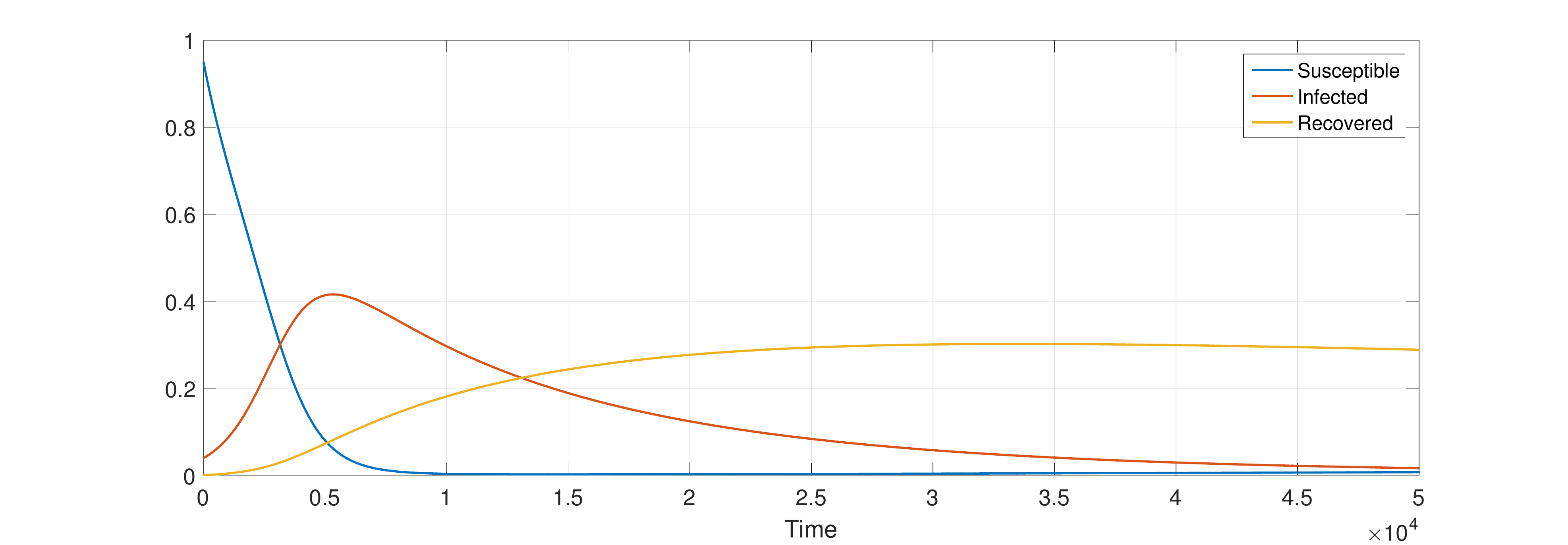}}
	\caption{Time variation of the obtained numerical solutions from (AP)-scheme with $\varepsilon=10^{-6}$ using initial condition $ii)$ and with diffusion, at $x = 0$, for the transmission rate values $\beta=0.1727,\,0.2763,\,0.449,\,0.6908,\,0.1.7269,\,5.1807$, the corresponding reproduction ratio is $R_0 = 0.5,\,0.8,\,1.3;\,2,\,5\,\,10,\,15$  }
	\label{F2}
\end{figure}

The aim is to illustrate the transmission rate function influence over the evolution of the pandemic. For this, we start by considering constant values of $\beta=0.1727,\,0.2763,\,0.449,\,0.6908,\,0.1.7269,\,5.1807$, (the corresponding reproduction ratio is $R_0 = 0.5,\,0.8,\,1.3;\,2,\,5\,\,10,\,15$  respectively). Figure \ref {F2} shows the variation over time of susceptible, infected and recovered individuals with diffusion case $b)$ at $ x = 0 $ performed with the initial condition $ i) $. It is clear that for low values of the transmission rate, the proportion of infected individuals is low. Moreover, the steady-state results in a relatively low proportion of the population among recovered individuals, while the majority of the population remains among susceptible individuals.
While, for relatively high and moderate values of $ \beta $, a large proportion of individuals is found in equilibrium among the individuals recovered.  In other words, most of the population caught the disease and got infected, and then recovered.
Note that, in this case, only a relatively small proportion of the population remains susceptible individuals. In addition, infected individuals disappear after a reasonable period of time, while susceptible and recovered individuals reach a non-zero constant at steady-state.

Now, let consider a time-dependent transmission rate $\beta(t)$ giving by the following step-wise function

\begin{equation}\label{bita}
	\beta(t)=0.1727\mathds{1}_{[0,T/4]}(t)+  1.1052\mathds{1}_{]T/4,T/2]}(t)+0.0691\mathds{1}_{]T/2,2T/4]}(t)+17.2691\mathds{1}_{]2T/4,T]}(t), 
\end{equation}   
where $T=10^{5}$.

\begin{figure}[h!]
	\centering
	\subfigure[]{\includegraphics[height=1.7in ,width=2in]{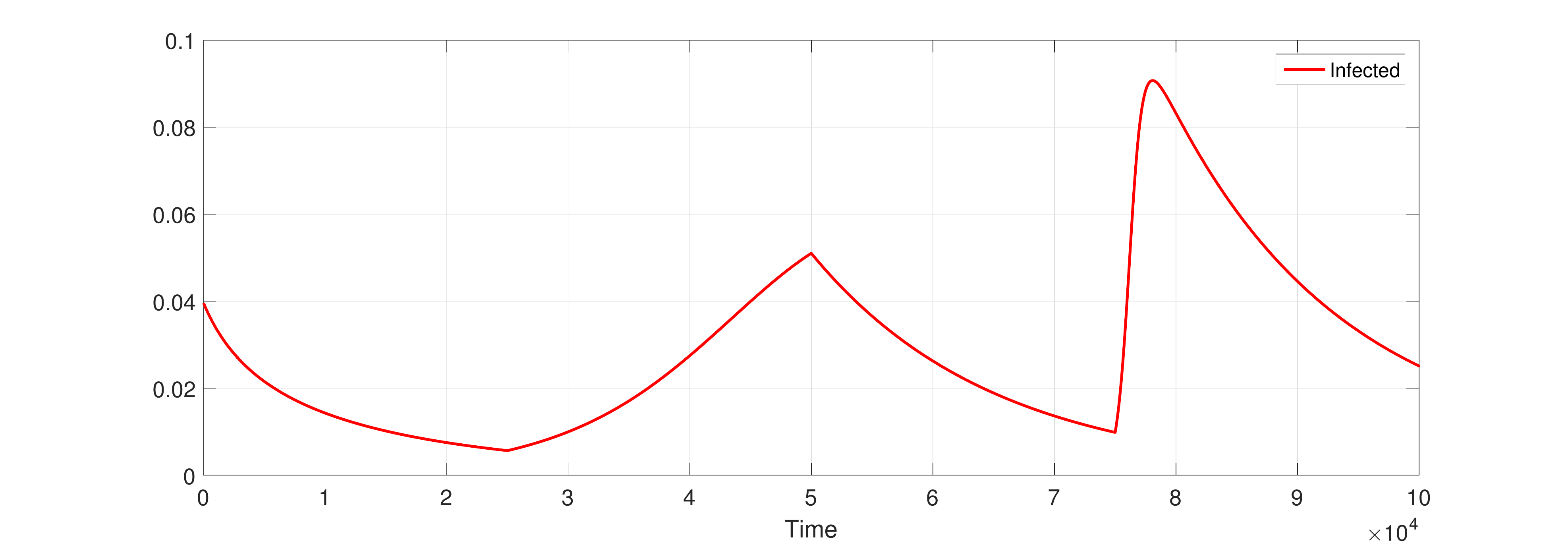}}~
	\subfigure[]{\includegraphics[height=1.7in ,width=2in]{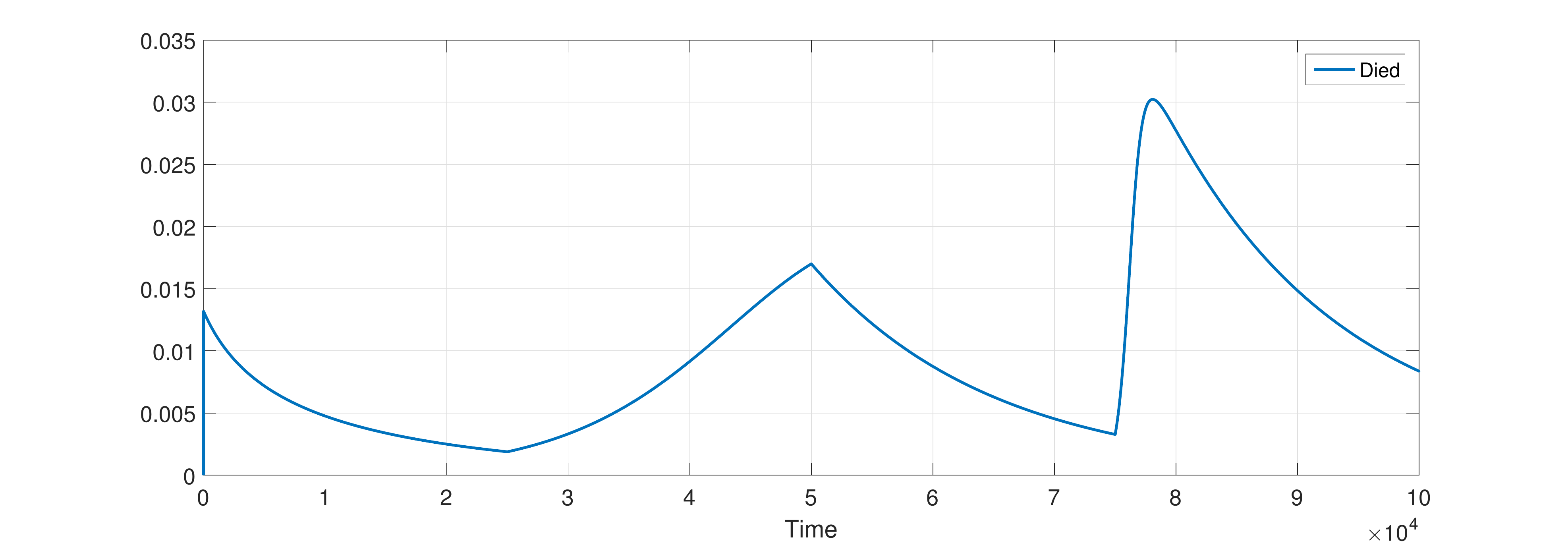}}
	\caption{Time variation of infected $a$ and died $b$ individuals obtained with asymptotic preserving numerical scheme using initial condition $ii)$ and diffusion case $b)$, at $x = 0$. The transmission rate function $\beta(t)$ is given by Eq. (\ref{bita}) and $\varepsilon=10^{-6}$.}
	\label{FB}
\end{figure}

Figure \ref{FB} presents time variation of infected and died individuals obtained with the asymptotic preserving numerical scheme with $\varepsilon=10^{-6}$, self-diffusion case $b)$ and initial condition $ii)$  at $x = 0$.  We observe  the numbers of infected and died individuals increase from the time $ T/4$ called the first wave, also at time $t>3T/4$ considered as the second wave occurs because of the values of  $\beta$ which corresponds  to $ R_0(t) >1 $. As time progresses, we notice that the numbers of infected and died individuals decrease at time $ t>T/2 $, the same at time $t<T/4$, thanks to the choice of  transmission rate function, where a small value of $ R_0(t) = 0.2 <1 $ and $ R_0(t) = 0.1 <1 $, respectively is considered.

\subsubsection{Test 3: self-diffusion effect by $\phi(x)$}
\begin{figure}[h!]
	\subfigure[]{\includegraphics[height=1.7in ,width=1.6in]{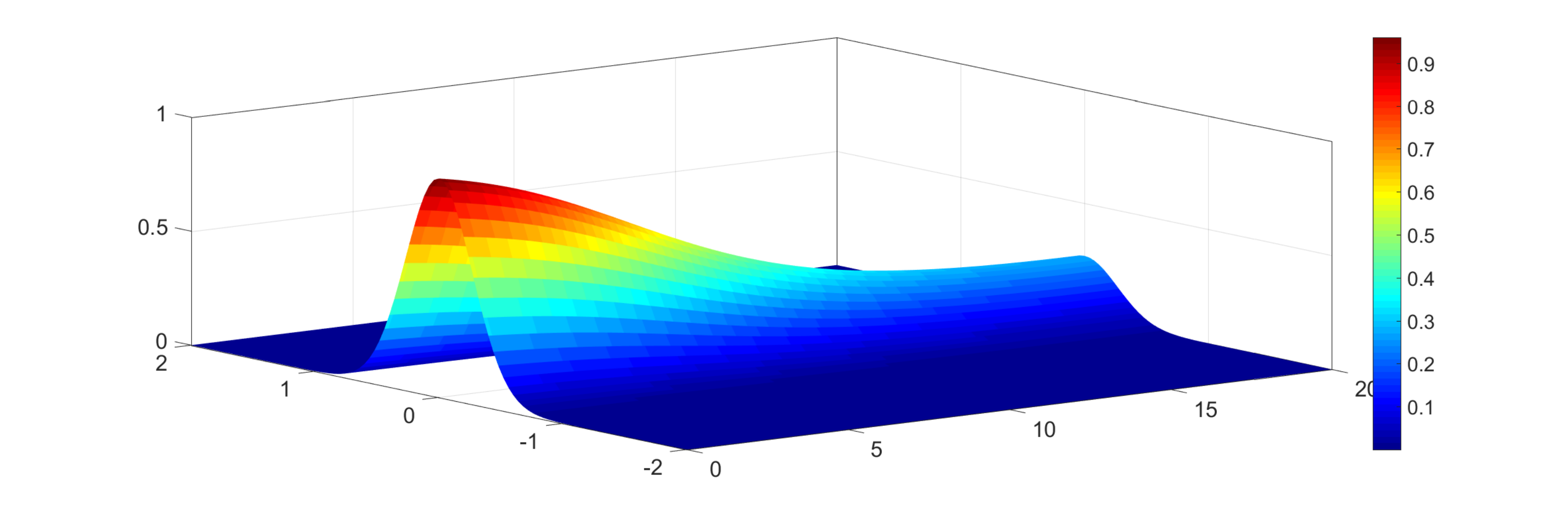}}~
	\subfigure[]{\includegraphics[height=1.7in ,width=1.6in]{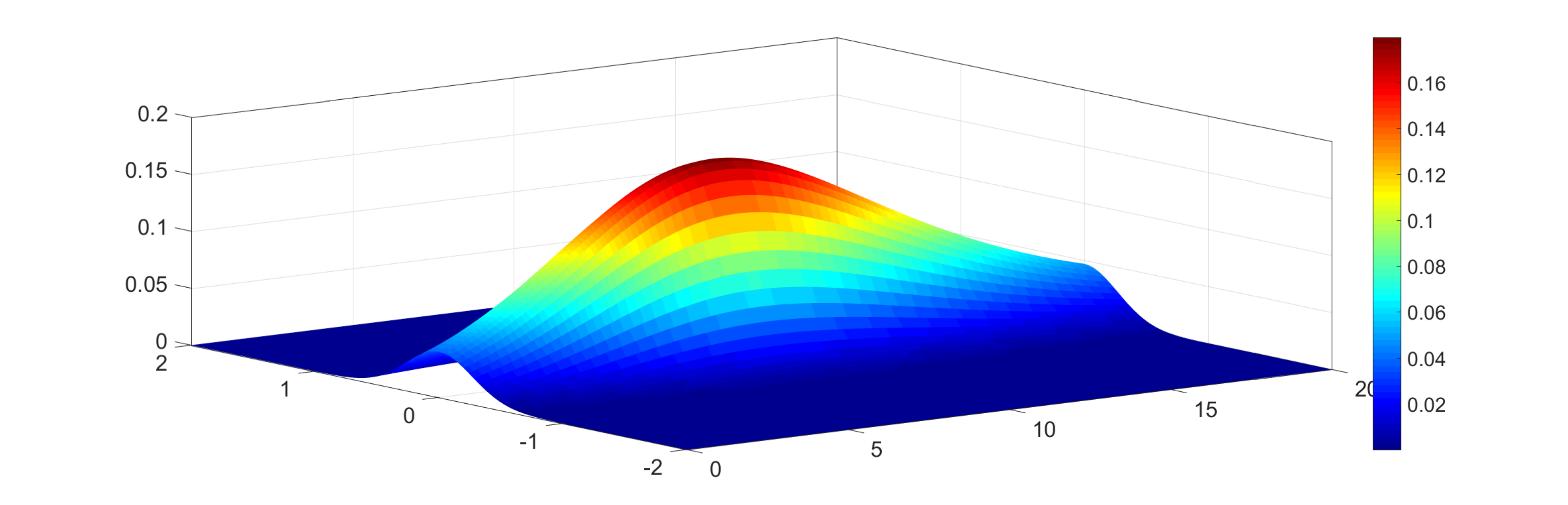}}~
	\subfigure[]{\includegraphics[height=1.7in ,width=1.6in]{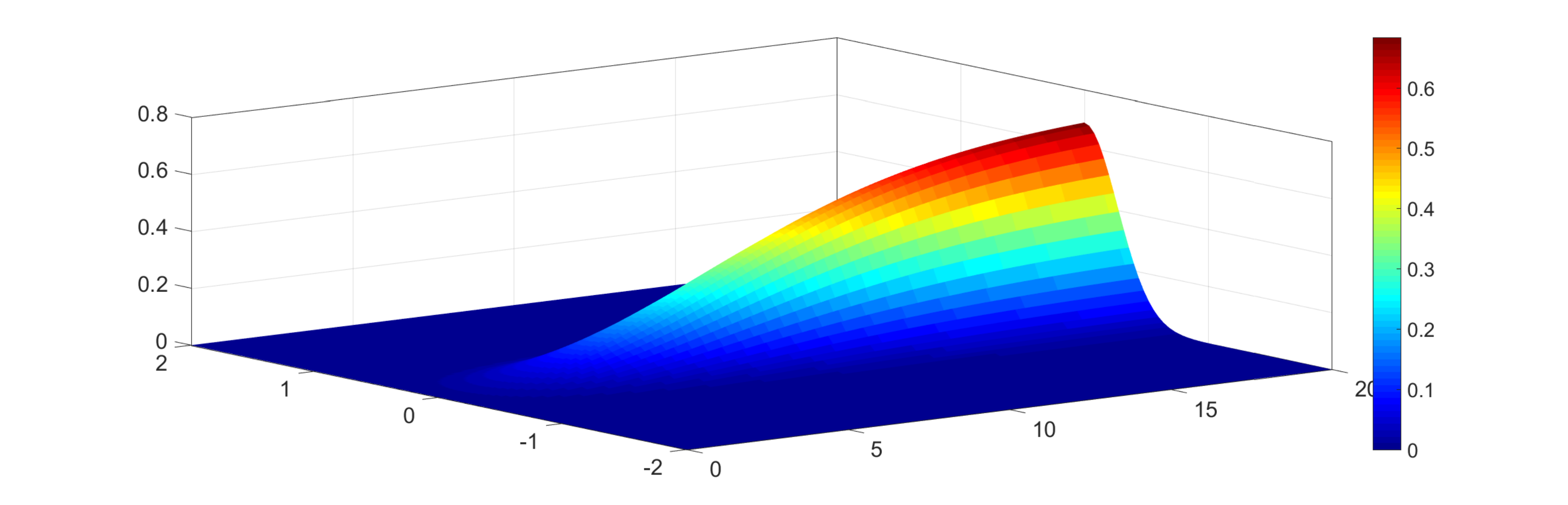}}	\subfigure[]{\includegraphics[height=1.7in ,width=1.6in]{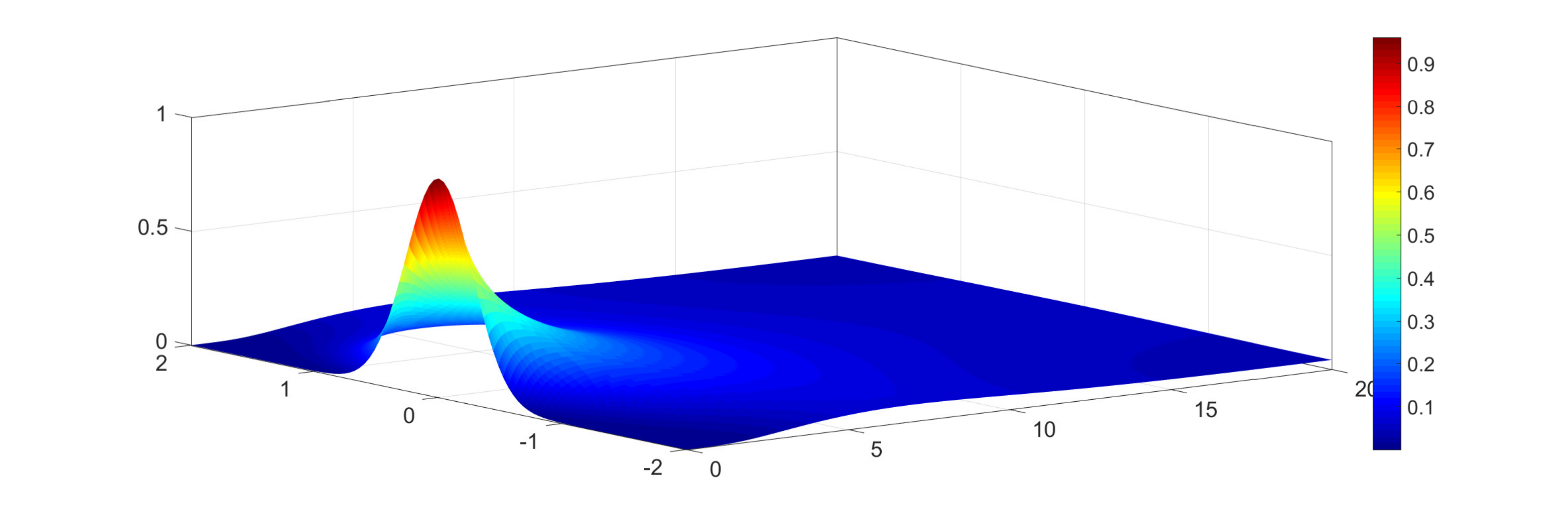}}~
	\subfigure[]{\includegraphics[height=1.7in ,width=1.6in]{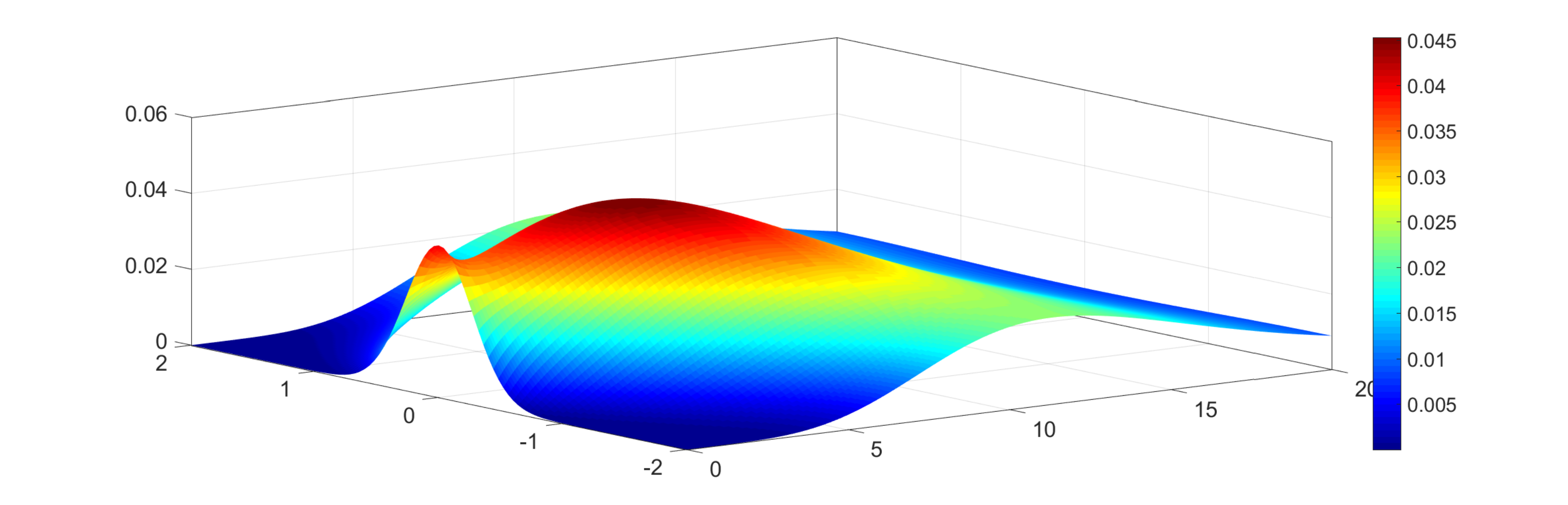}}~
	\subfigure[]{\includegraphics[height=1.7in ,width=1.6in]{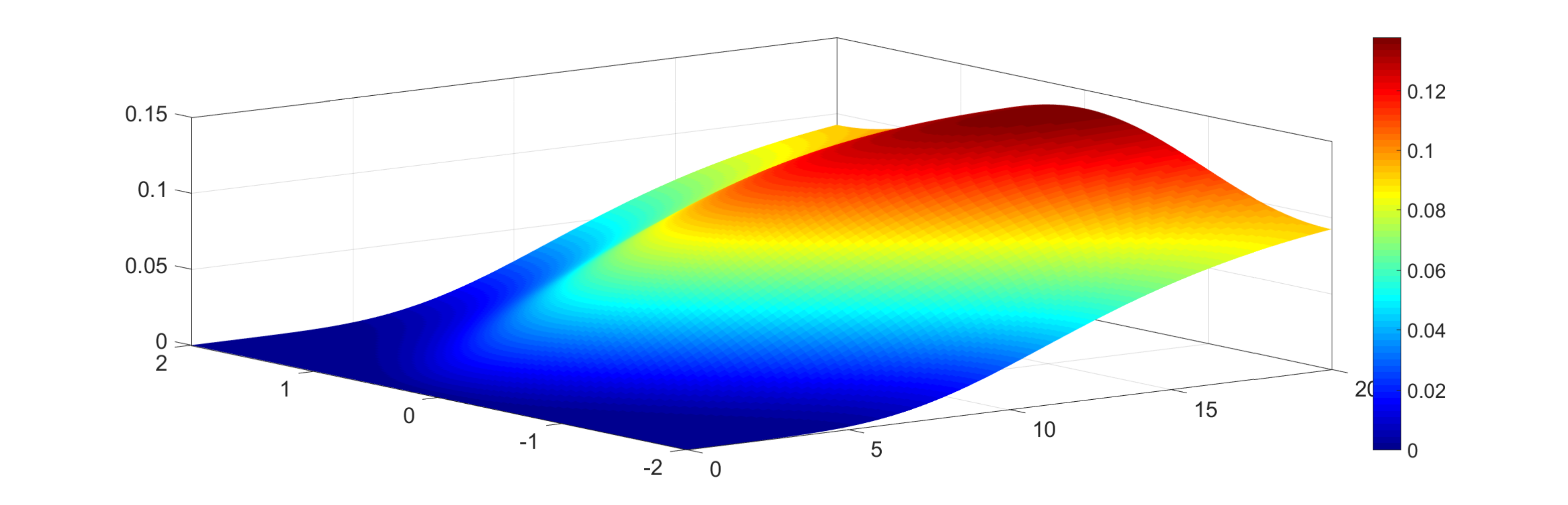}}
    \subfigure[]{\includegraphics[height=1.7in ,width=1.6in]{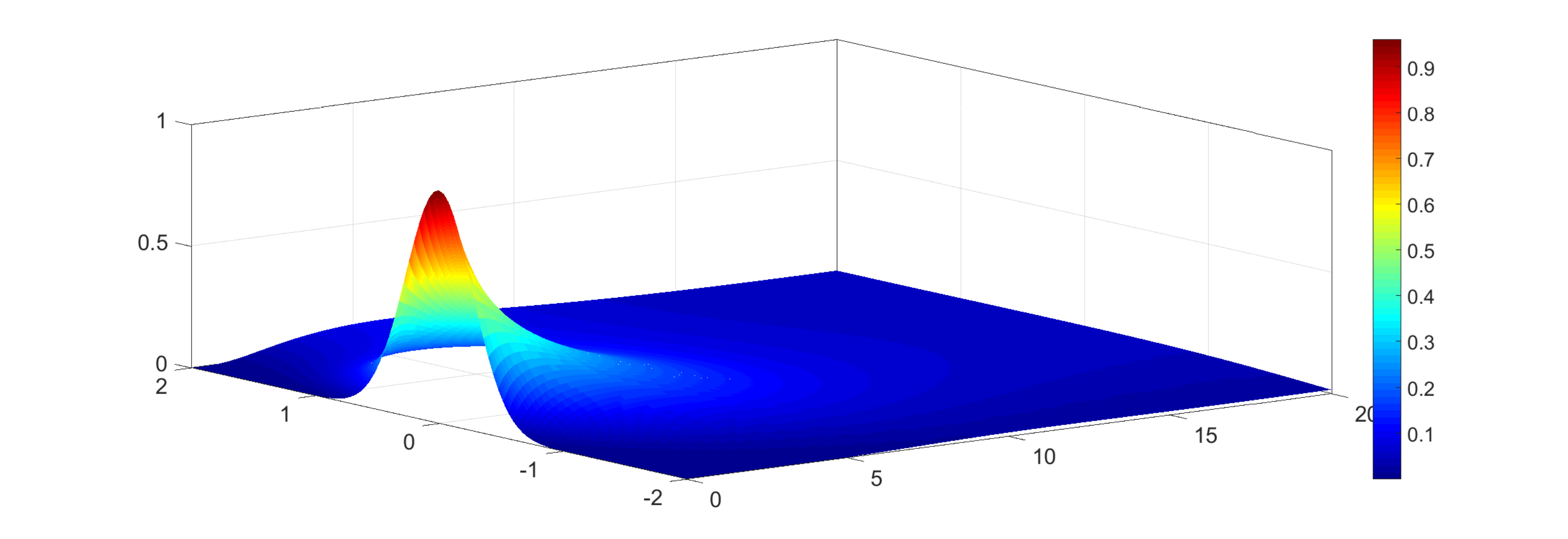}}~
    \subfigure[]{\includegraphics[height=1.7in ,width=1.6in]{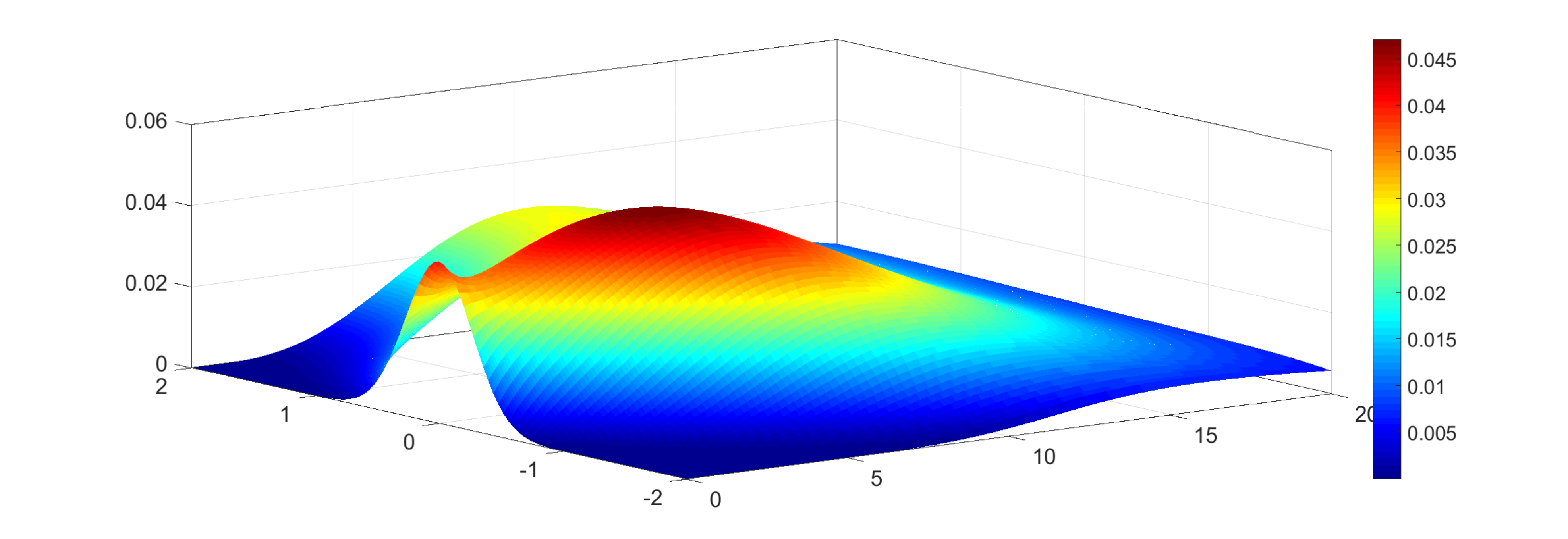}}~
     \subfigure[]{\includegraphics[height=1.7in ,width=1.6in]{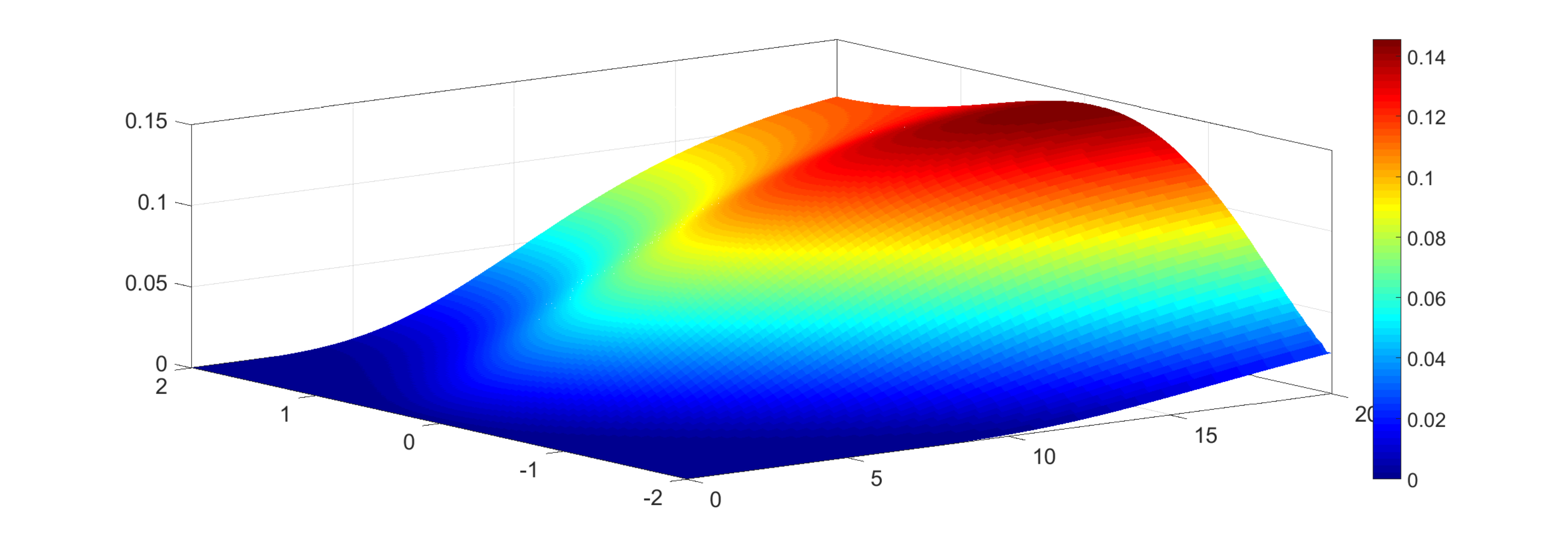}}
	\caption{Evolution of densities  $f_1$ (first column) $f_2$ (second column) and $f_3$ (third column) obtained with the asymptotic preserving numerical scheme for $\varepsilon=10^{-6}$ and initial condition $ii)$: without diffusion case $a)$ (first line), with diffusion case $b)$ (second line), and with diffusion case $c)$ (third line). The reproduction ratio is $R_0 = 2$.}
	\label{F3}
\end{figure}

This test shows the effect of self-diffusion over the interacting individuals. For this, let consider the initial conditions $ ii) $ and the reproduction ratio is $ R_0 = 2 $. In Figure \ref {F3}, we show the numerical results of susceptible, infected and recovered individuals obtained with asymptotic preserving scheme where $ \varepsilon = 10 ^ {- 6} $. Three cases are considered:  without diffusion, case $a)$ illustrated in sub-figures (a)-(b)-(c), with diffusion case $b)$  illustrated in sub-figures (d)-(e)-(f)), and with  diffusion case $c)$ illustrated in sub-figures (h)-(g)-(i). In the first case, the individuals are all centered around the axis $ x = 0$. In the second case where $\varphi_i=\mid x\mid$, we observe that individuals are more spreading within the domain. In the third case where $\varphi_i(x)=1+0.5\,x$, we notice that the individuals diffuse more on the positive $x$-axis.

\subsubsection{Test 4: cross-diffusion effect by $\chi(S,I)$}

In this test we show the effect of the cross-diffusion term over the interacting individuals. For this, we consider the initial conditions $ ii) $ and the reproduction ratio is $ R_0 = 2 $. 
Figure \ref {F3} illustrates the numerical results of susceptible, infected and recovered individuals obtained with asymptotic preserving scheme where $ \varepsilon = 10 ^ {- 6} $. Three cases are considered:  without cross-diffusion where $\chi=0$ illustrated in sub-figures (a)-(d), with cross-diffusion  where $\chi=0.01$ illustrated in sub-figures (d)-(e), and with  cross-diffusion where $\chi(S)=\frac{5S}{1+S^2}$ illustrated in sub-figures (c)-(f).

\begin{figure}[h!]
	\centering
	\subfigure[]{\includegraphics[height=1.7in ,width=2in]{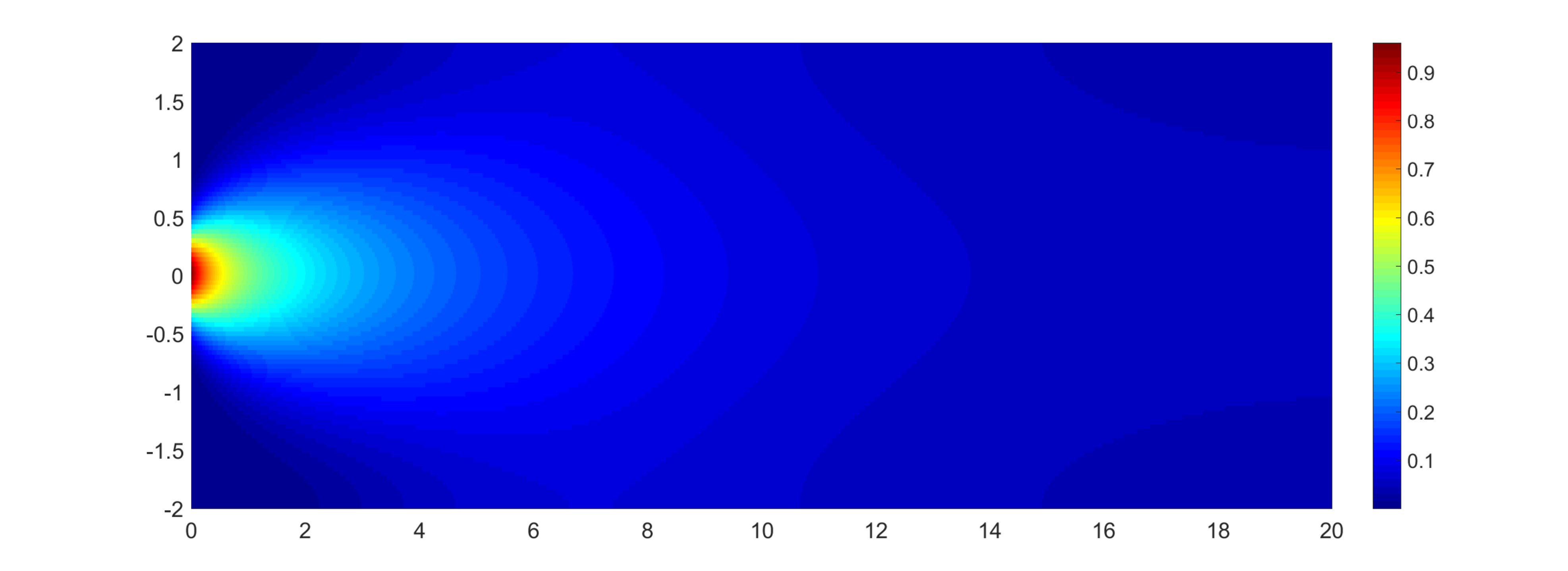}}~
	\subfigure[]{\includegraphics[height=1.7in ,width=2in]{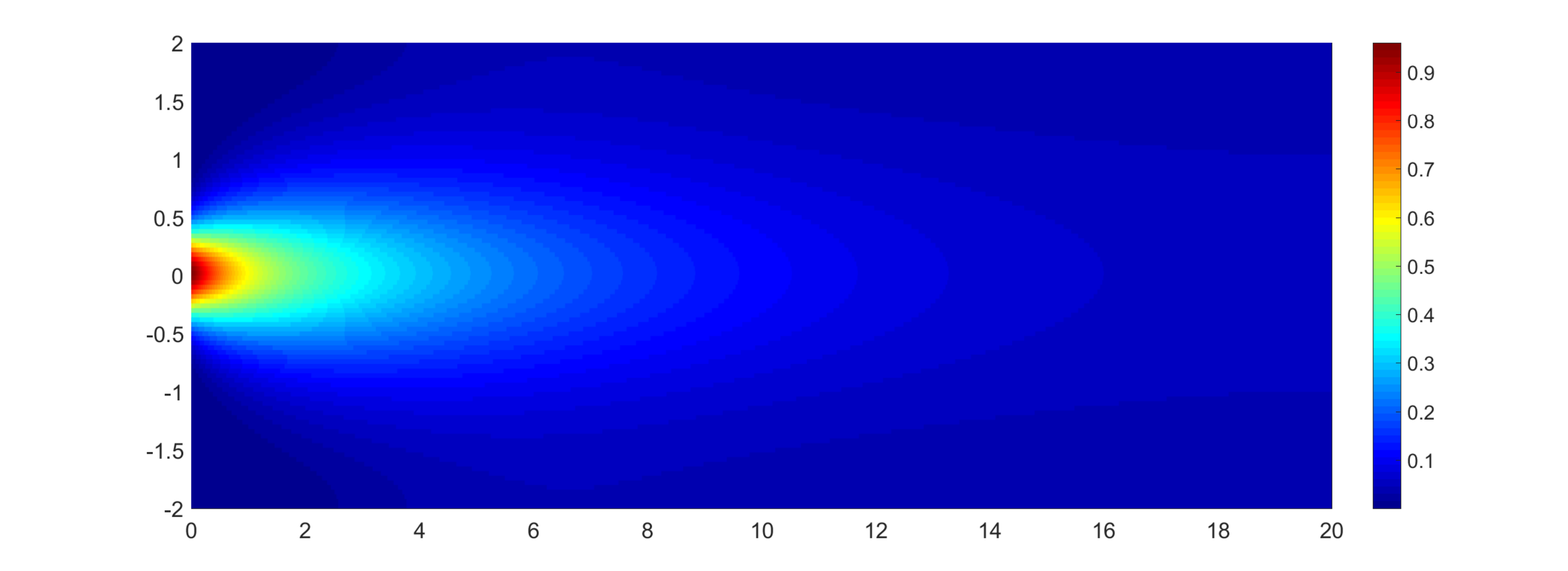}}~
	\subfigure[]{\includegraphics[height=1.7in ,width=2in]{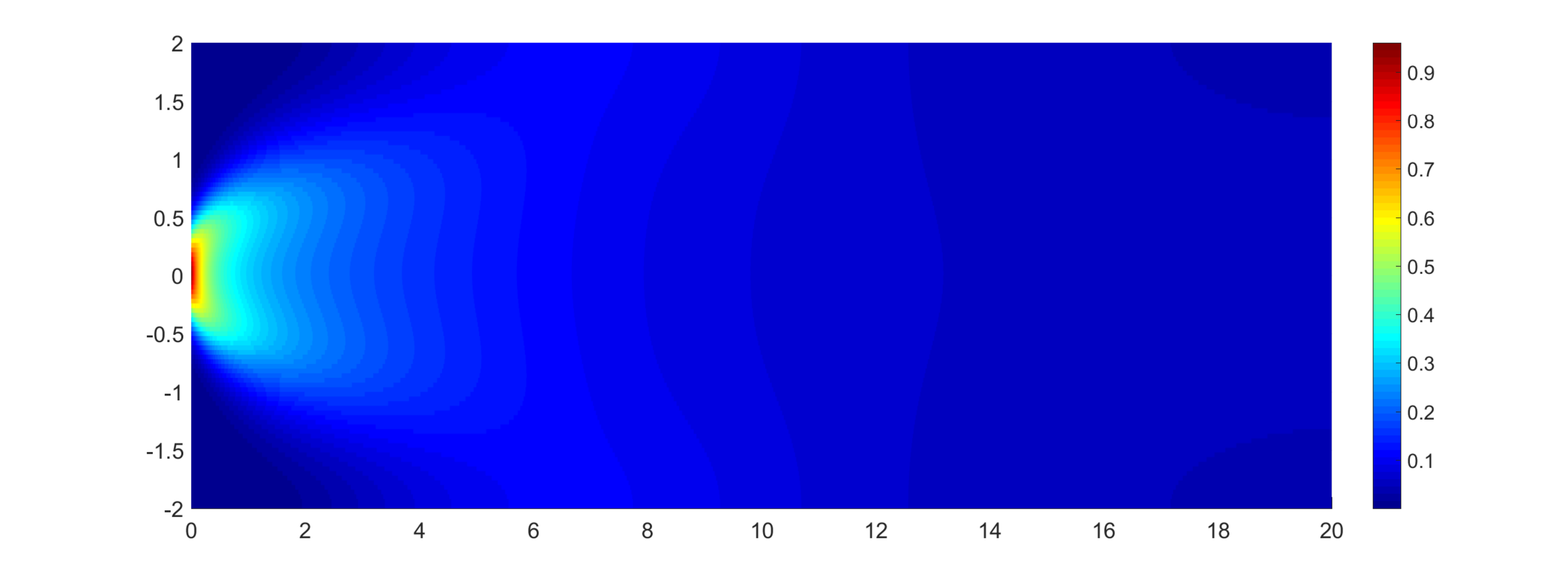}}
	\subfigure[]{\includegraphics[height=1.7in ,width=2in]{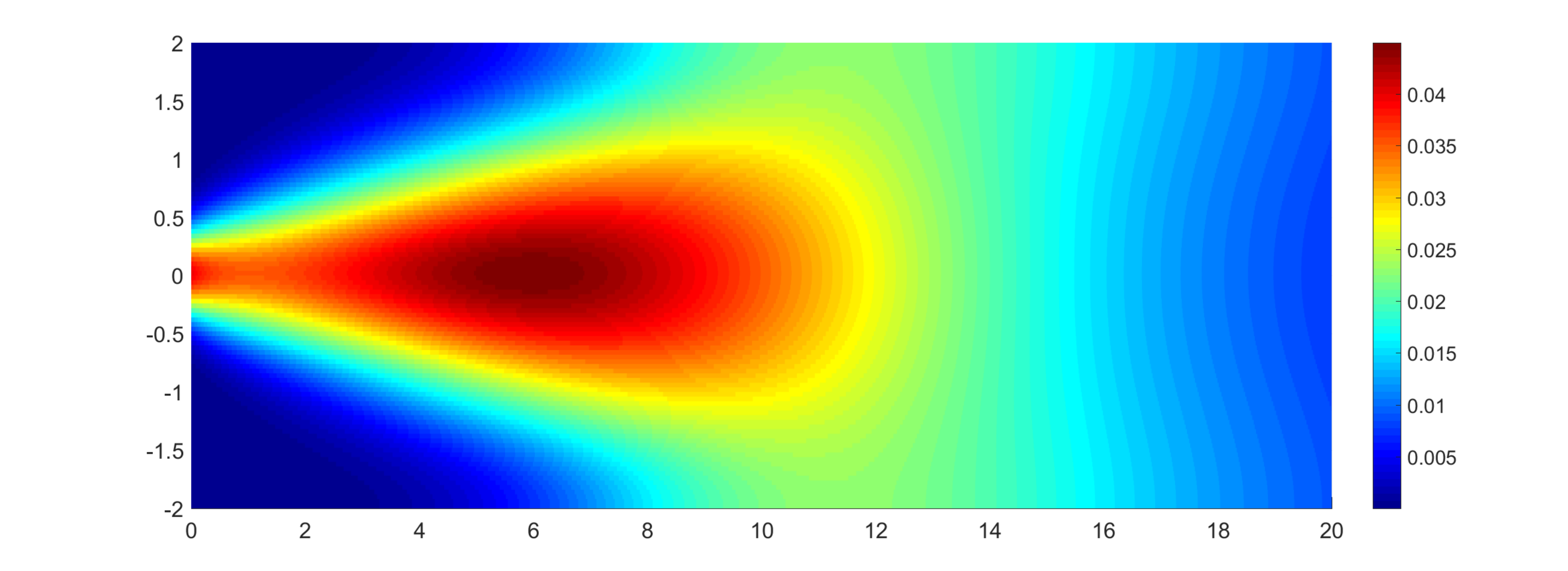}}~
	\subfigure[]{\includegraphics[height=1.7in ,width=2in]{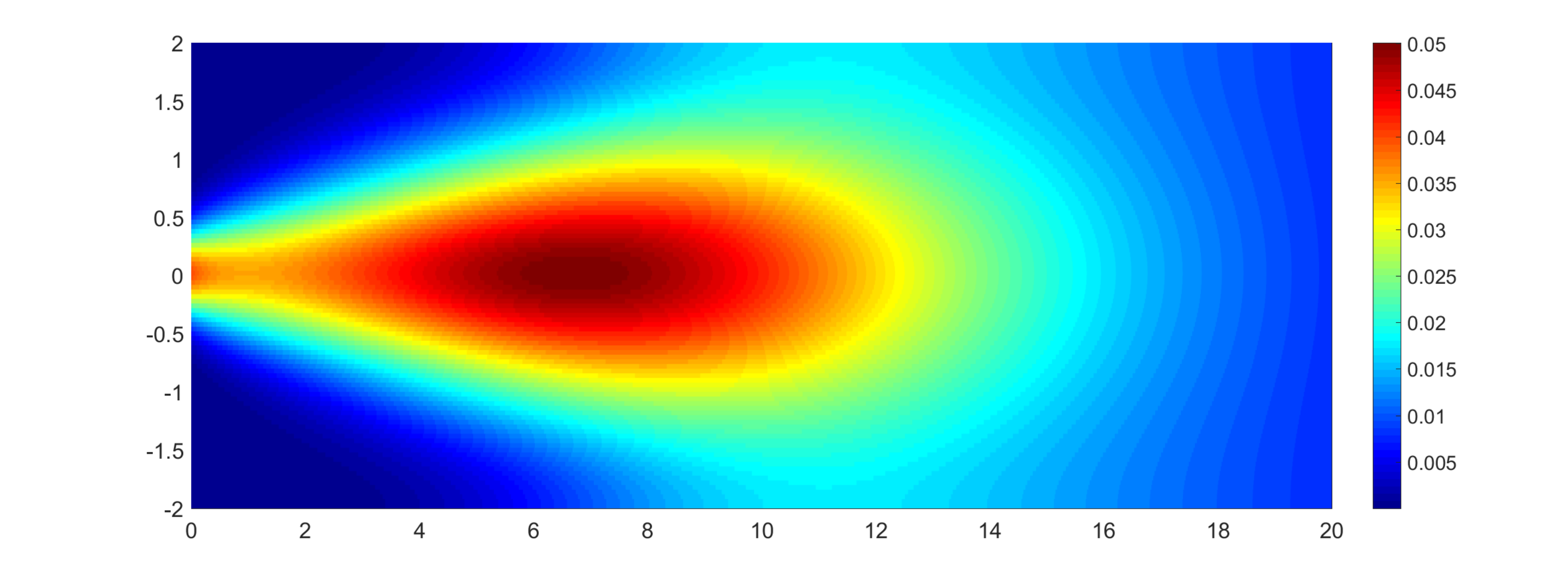}}~
	\subfigure[]{\includegraphics[height=1.7in ,width=2in]{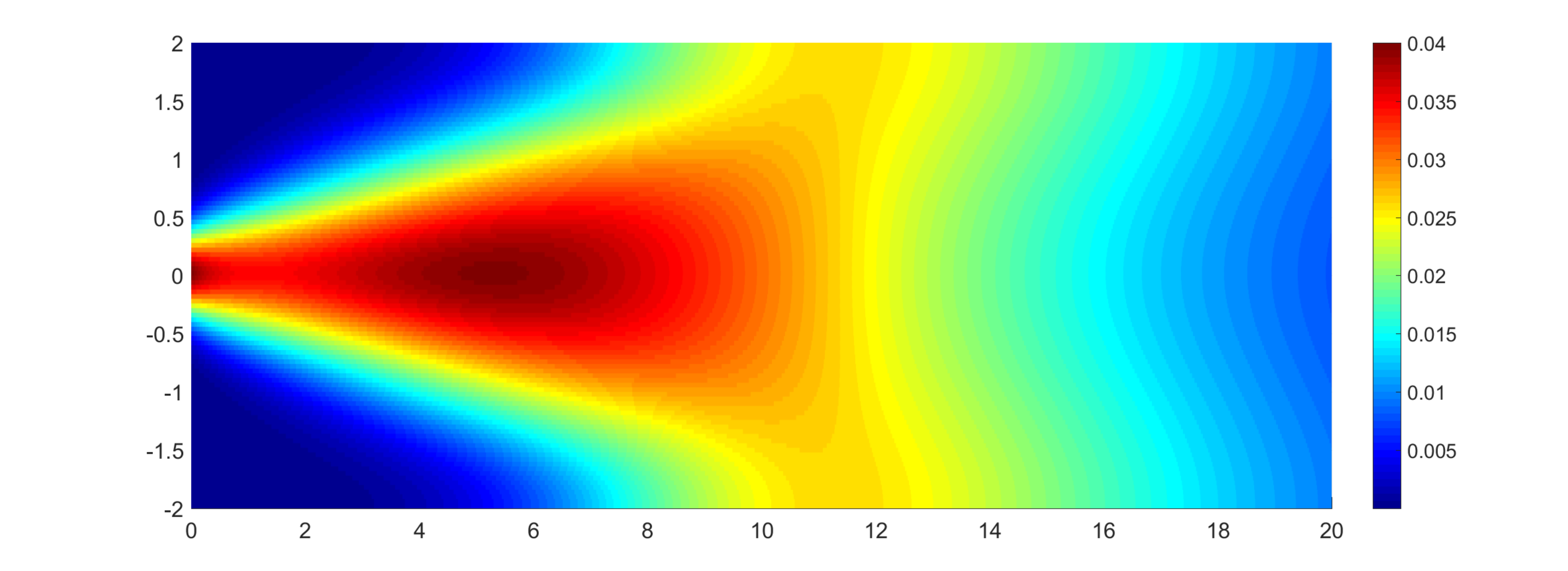}}
	\caption{Time variation of the obtained numerical solutions of Susceptible (first line) and Infected (second line) with (AP)-scheme with $\varepsilon=10^{-6}$ using initial condition $ii)$ and with diffusion, at $x = 0$, for different choices of $\chi(S,I)=0$ (left), $\chi(S,I)=0.01$ (middle), $\chi(S,I)=\frac{5S}{1+S^2}$. }
	\label{F4}
\end{figure}
\section{Computational analysis of SIRD cross-diffusion epidemic system in two dimensional space}  \label{Sec5}

Motivated by the numerical simulations in one dimension, we illustrate the behavior of time dependent nonlinear SIRD cross-diffusion epidemic system. Namely, we show the generated formation of patterns. The numerical investigation is performed using the finite volume method. 
\subsection{An implicit finite-volume scheme}
In order to solve numerically system \eqref{Cross-Diffusion}, we adopt the finite volume method in 2D. For that, we consider a family $\mathfrak{T}_h$ of admissible meshes of the domain $\Omega$ consisting of disjoint open and convex polygons called control volumes, see \cite{[EH00]}. In the rest of this subsection, we shall use the following notation: the parameter $h$ is the maximum diameter of the control volumes in $\mathfrak{T}_h$. $K$ is a generic volume in $\mathfrak{T}$, $|K|$is the $2$-dimensional Lebesgue measure of $K$ and $N(K)$ is the set of the neighbors of $K$. In addition, for all $L \in N(K)$, we denote by $\sigma_{K,L}$ the interface between $K$ and $L$ where $L$ is a generic neighbor of $K$. $\eta_{K,L}$ is the unit normal vector to $\sigma_{K,L}$ outward to $K$. For an interface $\sigma_{K,L}$, $|\sigma_{K,L}|$ will denote its $1$-dimensional measure. $d_{K,L}$ denotes the distance between $x_K$ and $x_L$, where the points $x_K$ and $x_L$ are respectively the center of $K$ and $L$.
We assume that a discrete function on the mesh $\mathfrak{T}_h$ is a set $(w_K)_K\in\mathfrak{T}$ and we identify it with the piece-wise constant function $w_h$ on $\Omega$ such that $w_h\mid_K = w_K$. Furthermore, we consider an admissible discretization of $(0,T)\times\Omega$ consisting of an admissible mesh $\mathfrak{T}_h$ of $\Omega$ and of a time step size $\Delta t_h > 0$ (both $\Delta t_h$ and the size $\max_{K\in t_h}diam(K)$ tend to zero as $h \to 0$). Now, let define the discrete gradient $\nabla_hw_h$ as the constant per diamond $T_{K,L}$ function by
\begin{equation*}\label {14}
\Big(\nabla_hw_h\Big)\rvert_{\mathfrak{T}_{K,L}}=\nabla_{K,L}w_h:=\frac{w_L-w_K}{d_{K,L}}\eta_{K,L}.
\end{equation*}
Finally, we define the average of source terms $F_{i,K}^{n+1}$ by $F_{i,K}^{n+1}=F_i(S(t^n,x),I(t^n,x),R(t^n,x)),$ for $i=1,2,3$. And we make the following choice to approximate the function  $\chi_{K,L}^{n+1}$
$$\chi_{K,L}^{n+1}=\chi\big(\min\{S_{K}^{{n+1}^+},S_{L}^{{n+1}^+}\},\min\{I_{K}^{{n+1}^+},I_{L}^{{n+1}^+}\}\big),$$
where $u_{i,J}^{{n+1}^+}=\max(0,u_{i,J}^{n+1})$ for $i=1,2,3$ and $J=K,L$.
The computation starts from the initial cell averages $\displaystyle u_{i,0}^K=\frac{1}{|K|}\int_{K}u_{i,0}(x)\,dx$ for $i=1,2,3$.
To advance the numerical solution from $t^n$ to $t^{n+1} = t^n + \Delta t$, we use the following implicit finite
volume scheme: determine $S^{n+1}_{K},$ $I^{n+1}_{K}$, $R^{n+1}_{K}$ and $D^{n+1}_{K}$ for $K\in \mathfrak{T}$ such that
\begingroup\small
\begin{equation}\label{2DScheme}
\left\{
\begin{array}{ll}
|K|\frac{S^{n+1}_{K}-S^{n}_{K}}{\Delta t}-d_{1}\sum\limits_{L \in N(K)}	\frac{|\sigma_{K,L}|}{d_{K,L}}(S_{L}^{n+1}-S_{K}^{n+1})\\
\hskip.2cm+\sum\limits_{L \in N(K)}\frac{|\sigma_{K,L}|} {d_{K,L}}\Big[\chi_{K,L}^{n+1}(S_{L}^{n+1}-S_{K}^{n+1})+\chi_{K,L}^{n+1}(I_{L}^{n+1}-I_{K}^{n+1})\Big] =|K| F_{1,K}^{n},\\
|K|\frac{I^{n+1}_{K}-I^{n}_{K}}{\Delta t}
-d_{2}\sum\limits_{L \in N(K)}	\frac{|\sigma_{K,L}|}{d_{K,L}}(I_{L}^{n+1}-I_{K}^{n+1})=|K| F_{2,K}^{n}, \\
|K|\frac{R^{n+1}_{K}-R^{n}_{K}}{\Delta t}-d_{3}\sum\limits_{L \in N(K)}\frac{|\sigma_{K,L}|}{d_{K,L}}(R_{L}^{n+1}-R_{K}^{n+1})=|K| F_{3,K}^{n},\\
|K|\frac{D^{n+1}_{K}-D^{n}_{K}}{\Delta t}=\alpha I_{K}^{n+1},
\end{array}\right.
\end{equation}\endgroup
for all $K\in \mathfrak{T}_h,\; n\in N_h$.  We consider implicitly the homogeneous Neumann boundary condition and Newton method has been used in order to solve the corresponding nonlinear system arising from the implicit finite volume scheme \eqref{2DScheme}. Note that the linear systems involved in Newton's method are solved by the GMRES method.

\subsection{Numerical simulations}
The numerical simulations are performed by uniform mesh given by a Cartesian grid $N_x=N_y=200$ in the space domain $\Omega=(0,0.5)\times(0,0.5)$. The time stepping is explicit with a fixed time step $\Delta t=0.001$. The model parameters are set to $\mu=1/83,\,\gamma=1/3,\,R_0=5$, the constant self coefficients are chosen to be $d_1=0.025,\,d_2=0.015,\,d_3=0.001$, and cross-diffusion term is given by $\chi(S)=\frac{5S}{1+S^2}$. We mention that the patterns of the species $S$ coincide with those of $I$,
therefore they are not shown.

\subsubsection{Example 1} We assume that the density of sub populations is a random perturbation around the endemic stationary state $(S^*,I^*,R^*)$. Thus, the initial data are given by
$$S(0,x)=S^*+S(x)_\delta,\quad I(0,x)=I^*+I(x)_\delta,\quad R(0,x)=R^*+R(x)_\delta,\qquad x\in\Omega,$$
where $J(x)_\delta\in[0,1]$ is a uniform distributed variable for $J=S,\,I,\,R$. The stationary state is given by 

$$(S^*,I^*,R^*)=\big((\gamma+\mu)/\beta,\mu(R_0-1)/\beta,\gamma(R_0-1)/\beta\big).$$
  
\begin{figure}[h!]
	\centering
	\subfigure[]{\includegraphics[height=1.7in ,width=2in]{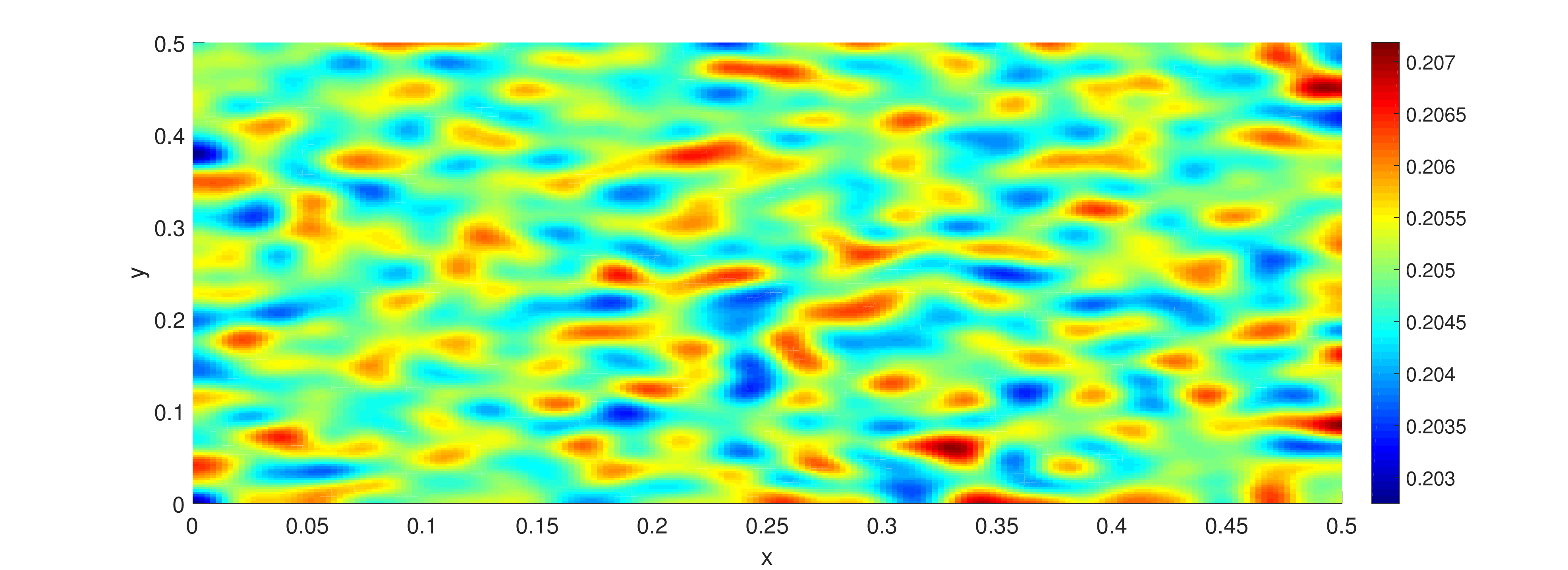}}~
	\subfigure[]{\includegraphics[height=1.7in ,width=2in]{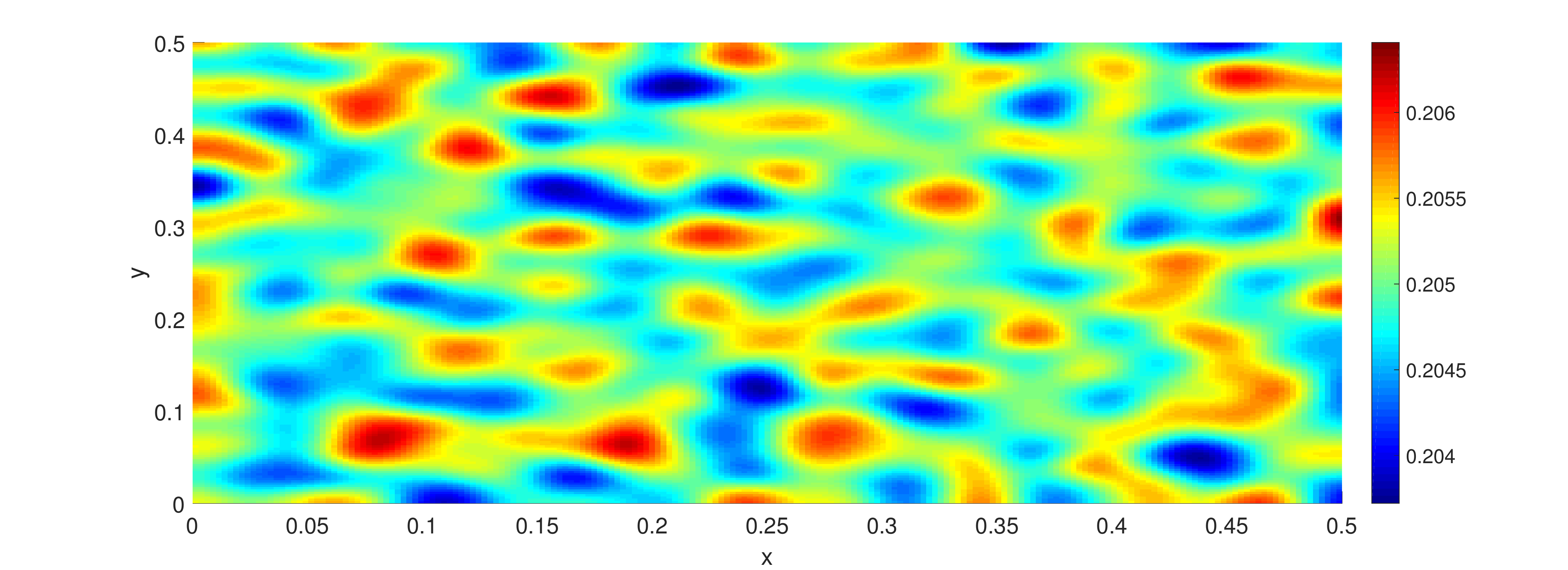}}~
	\subfigure[]{\includegraphics[height=1.7in ,width=2in]{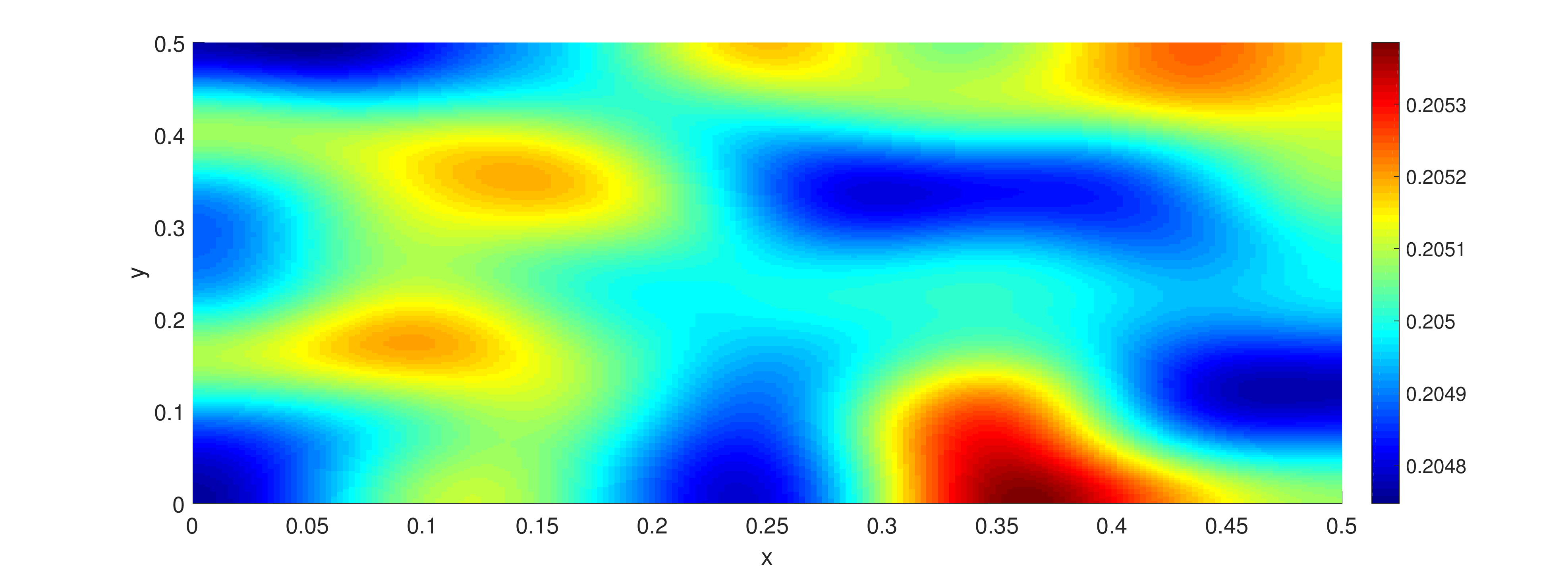}}
	\caption{Numerical solution for $S$ at time instants $t = 0.005, t = 0.01$ and $t = 0.1$ (Example 1).}
	\label{F5}
\end{figure}

In Figure \ref{F5}, we observe islands of high concentration of susceptible individuals are formed. In fact, this reflects the phase separation triggered by the susceptible subpopulation avoiding the infected subpopulation.

\subsubsection{Example 2} For this Example, the only difference from Example 1 is that the initial data is now randomly distributed at only four spatial points as follows
$$S(0,x)=S^*+\sum_{i=1}^{4}S(x_i)_\delta,\quad I(0,x)=I^*+\sum_{i=1}^{4}I(x_i)_\delta,\quad R(0,x)=R^*+\sum_{i=1}^{4}R(x_i)_\delta,\qquad x\in\Omega,$$
where $x_1=(1/8,1/8),\;x_2=(3/8,1/8),\;x_3=(1/8,3/8),\;x_4=(3/8,3/8)$.

\begin{figure}[h!]
	\centering
	\subfigure[]{\includegraphics[height=1.7in ,width=2in]{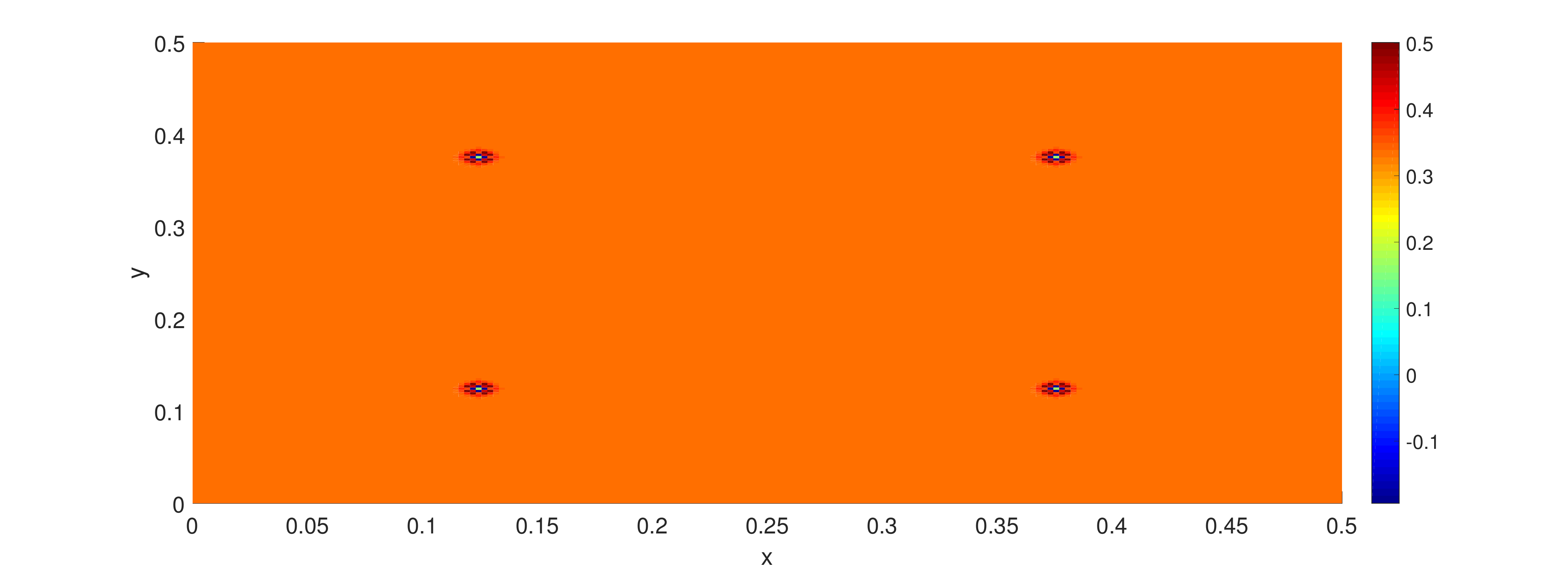}}~
	\subfigure[]{\includegraphics[height=1.7in ,width=2in]{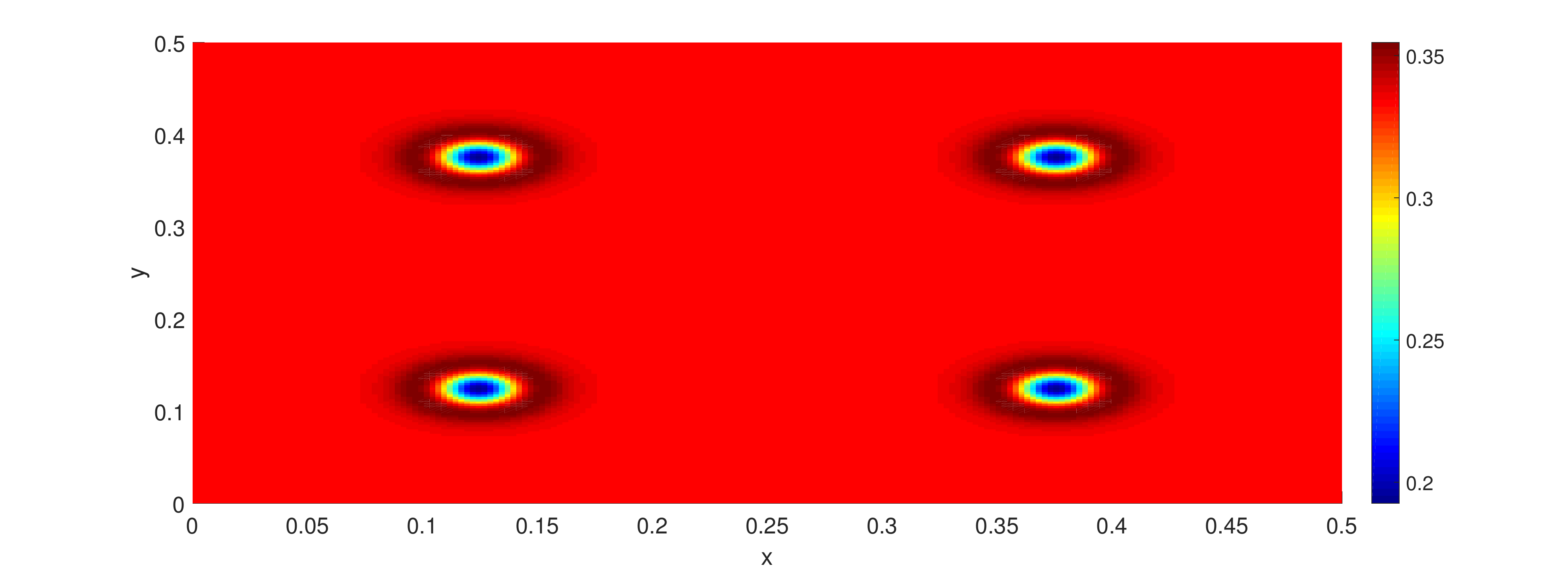}}~
	\subfigure[]{\includegraphics[height=1.7in ,width=2in]{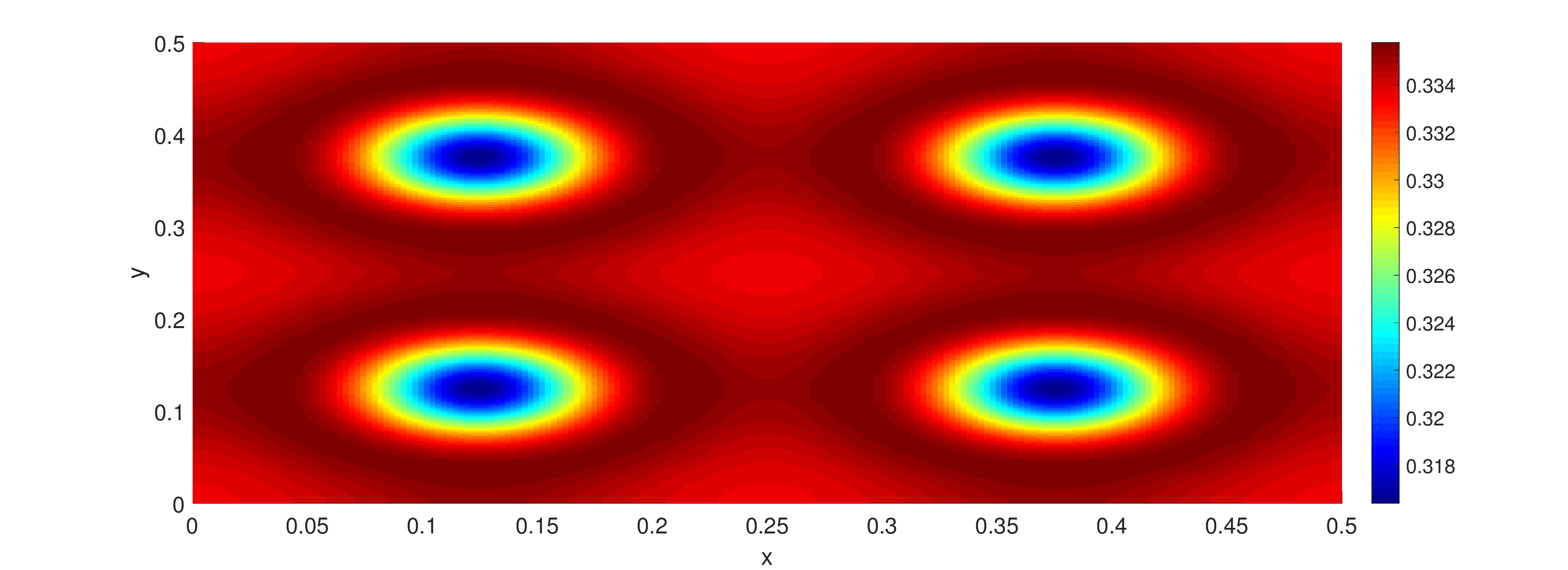}}
	\caption{Numerical solution for $S$ at time instants $t = 0.0001, t = 0.005$ and $t = 0.05$ (Example 2).}
	\label{F6}
\end{figure}

In Figure \ref{F6}, we notice that the perturbation in four single point leads to pattern formation in the whole domain and the spatial patterns become clearly visible at earlier time steps.
\section{Conclusion and perspectives}

In this paper, a time-independent SIRD nonlinear cross-diffusion system for epidemic has been proposed and derived from a kinetic theory model by using multiscale approach. Several numerical simulations have been provided. Specifically, the uniform stability along the transition from kinetic to macroscopic regimes is shown and the sensitivity to the transmission rate is demonstrated where the epidemic waves are depicted. Moreover, it has shown that the presence of the self and cross-diffusion terms in system \eqref{Cross-Diffusion} influences the spreading of the pandemic. In addition, we provided numerical simulations in two dimensional space where the generated formation of patterns are presented in two examples.\\ 
We believe that this paper opens such interesting perspectives: For instance, extension of the proposed macroscopic model by considering a time-space diffusion $d_i(t,x)$ and the rate transmission $\beta(t,x)$.

\end{document}